\RequirePackage[2020-02-02]{latexrelease}
\documentclass[column,preprint,numbers,amsmath,amssymb]{revtex4}
\usepackage{graphicx}
\usepackage{multirow}
\usepackage{amsmath}
\usepackage{dcolumn}
\usepackage{bm}
\usepackage{float}
\begin{document}
\title{Anisotropic Electron-Hole Excitation and Large Linear Dichroism in Two-Dimensional Ferromagnet CrSBr with In-Plane Magnetization}
\author{Tian-Xiang Qian,$^{1,2}$ Ju Zhou,$^{1,3}$ Tian-Yi Cai,$^{1,}$\footnote{Electronic address:
		caitianyi@suda.edu.cn} and Sheng Ju$^{1,}$\footnote{Electronic address: jusheng@suda.edu.cn}} \affiliation{$^{1}$ Institute of Theoretical and Applied Physics, School of Physical Science and Technology and Jiangsu Key Laboratory of Thin Film, Soochow University, Suzhou
	215006, P. R. China \\ $^{2}$ Dipartimento di Fisica, Università degli Studi di Milano, via Celoria 16, I-20133 Milano, Italy \\ $^{3}$ School of Mathematics and Physics, Queen's University Belfast, Belfast BT7 1NN, Northern Ireland, United Kingdom}
\begin{abstract}
{The observation of magnetic ordering in
atomically thin CrI$_3$ and Cr$_2$Ge$_2$Te$_6$ monolayers has aroused intense
interest in condensed matter physics and material science. Studies of van de Waals two-dimensional (2D) magnetic materials are of both fundamental importance and application interest. In particular, exciton-enhanced magneto-optical properties revealed in CrI$_3$ and CrBr$_3$ monolayers have expanded the understanding of exciton physics in 2D materials. Unlike CrI$_3$ and CrBr$_3$ with out-of-plane magnetization, CrSBr has an in-plane magnetic moment, therefore, providing a good opportunity to study the magnetic linear dichroism and high-order magneto-optical effects. Here, based on the many-body perturbation method within density-functional theory, we have studied quasiparticle electronic structure, exciton, and optical properties in CrSBr monolayer. Strongly bounded exciton has been identified with the first bright exciton located at 1.35 eV, in good agreement with an experiment of photoluminescence (Nat. Mater. \textbf{20}, 1657 (2021)). Strong contrast in the optical absorption is found between the electric fields lying along the in-plane two orthogonal directions. In accordance with a typical and realistic experimental setup, we show that the rotation angle of linear polarized light, either reflected or transmitted, could be comparable with those revealed in black phosphorene. Such large linear dichroism arises mainly from anisotropic in-plane crystal structure. The magnetic contribution from the off-diagonal component of dielectric function to the linear dichroism in CrSBr is negligible. Our findings not only have revealed excitonic effect on the optical and magneto-optical properties in 2D ferromagnet CrSBr, but also have shown its potential applications in 2D optics and optoelectronics.}
\end{abstract}

\maketitle

\section{introduction}

Due to strong light-matter interaction, two-dimensional (2D) materials have demonstrated potential applications in semiconductor optoelectronics and photonics \cite{xia,mak,sun,xiao}. The discovery of 2D van der Waals magnets Cr$_2$Ge$_2$Te$_6$ \cite{zhang} and CrI$_3$ \cite{xu} monolayer, on the other hand, has provided an additional degree of freedom, where the out-of-plane magnetization could facilitate magneto-optical Kerr and Faraday effects in ultrathin limit \cite{guo1,guo2}. Based on many-body perturbation theory, Wu, Cao, Li, and Louie have revealed exciton effect on the magneto-optical effects in the monolayer of CrI$_3$ \cite{wu-natcomm} and CrBr$_3$ \cite{wu-prmater}. The Kerr rotation angle for the reflected light could reach as high as 0.9$^{\circ}$ and the Fraday angle for the transmitted light could reach as high as 0.3$^{\circ}$. Recently, another kind of 2D magnet CrSBr with in-plane magnetization has been identified \cite{maz} and fabricated \cite{wilson} successfully. The bilayer system shows great contrast in the optical responses between antiferromagnetic and ferromagnetic interlayer coupling \cite{wilson}. Considering the excitonic effect, such a magnetic ordering dependence of optical properties has been well explained by the many-body perturbation calculations \cite{wilson}. In monolayer limit, the system shows strong photoluminescence (PL) at 1.3 eV \cite{wilson}. In the meantime, the exciton-coupled coherent magnons will lead to efficient optical access to spin information \cite{xyzhu}. In contrast to Cr$_2$Ge$_2$Te$_6$ and CrI$_3$ (CrBr$_3$), when the magnetization in CrSBr monolayer is lying along the easy axis within the 2D plane, as demonstrated in Fig. 1, the former magneto-optical Kerr and Fraday effects, which measures rotation angle of linear polarized light, will be replaced by Sch\"{a}fer-Hubert (SH) effect and Voigt effect, for reflected and transmitted lights, respectively \cite{tesa}. The original purpose of this paper is to provide a comprehensive framework of excitonic effect on the optical and high-order magneto-optical properties in 2D ferromagnetic CrSBr with in-plane magnetization. Generally, when the magnetization is pointing along $y$ direction, it is believed that \cite{tesa}

\begin{figure}
	\includegraphics[width=0.5\textwidth]{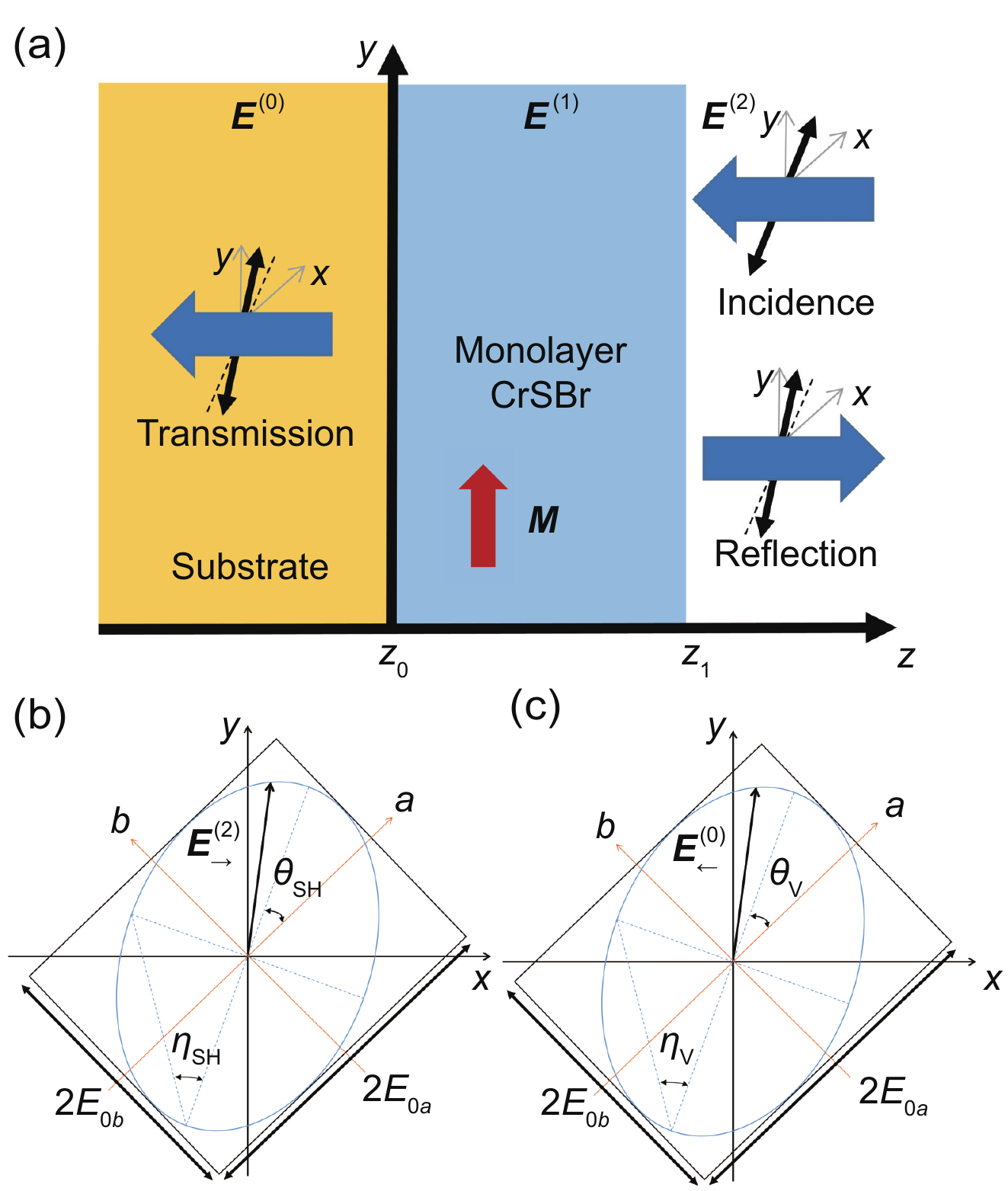}
	\caption{\label{fig:1} SH and Vogit setup consisting of layers of vacuum, ferromagnetic monolayer CrSBr, substrate of semi-infinitely thick SiO$_2$. Red arrow denotes the in-plane magnetization, which is along \textit{y} axis. (b) Illustration of rotation angle and ellipticity for reflected (left panel) and transmitted (right panel) lights.}
\end{figure}

\begin{equation}
\begin{aligned}
\phi_{\rm SH} 
& = \theta_{\rm SH} + i\eta_{\rm SH} 
& = \frac{-i\omega d}{c(1-n_{sub}^{2})}(n_{\parallel}^{2}-n_{\perp}^{2}) 
& = \frac{-i\omega d}{c(1-n_{sub}^{2})}(\varepsilon_{yy} - \varepsilon_{xx} - \frac{\varepsilon_{zx}^2}{\varepsilon_{zz}}),
\end{aligned}
\end{equation} 

and

\begin{equation}
\begin{aligned}
\phi_{\rm V} 
& = \theta_{\rm V} - i\eta_{\rm V} 
& = \frac{\omega d}{2ic}(n_{\parallel}-n_{\perp}) 
& = \frac{\omega d}{2ic}[\varepsilon_{yy}^{\frac{1}{2}} - (\varepsilon_{xx} + \frac{\varepsilon_{zx}^2}{\varepsilon_{zz}})^{\frac{1}{2}}].
\end{aligned}
\end{equation}
Here, $\theta_{\rm SH}$ and $\theta_{\rm V}$ are rotation angles, and $\eta_{\rm SH}$ and $\eta_{\rm V}$ are ellipticities. $c$ is the velocity of light, $n$ is the complex refraction index, and $\varepsilon$ is the dielectric function. $\omega$ is the optical frequency, $d$ is the thickness of magnetic material, and $sub$ means the substrate. Clearly, such effects are closely related to the contrast between the dielectric properties of two in-plane diagonal components $\varepsilon_{xx}$ and $\varepsilon_{yy}$ as well as the magnitude of the off-diagonal component $\varepsilon_{zx}$ of the system of interest. Based on independent particle approximation (IPA), such effects in 2D magnets CrXY (X=S, Se, and Te; Y=Cl, Br, and I) were studied computationally \cite{yao}. For the monolayer geometry with particular 2D dielectric screening, the inherent many-body correction to the quasiparticle band structure and optical properties could not be ignored \cite{book-louie,kel,rub,hyb,rod,thy}. A huge excitonic effect with binding energy of as large as 1 eV (two orders larger than conventional semiconductors) will modify the optical spectrum from IPA strongly. Therefore, in this paper, based on density functional theory with many-body perturbation method, i.e., \textit{GW}-BSE method (\textit{G} for one-particle Green's function, \textit{W} for screened Coulomb interaction, and BSE for Bethe-Salpeter equation) \cite{reinning}, we have studied the quasiparticle electronic structure, exciton, and optical properties in CrSBr monolayer. The first bright excitonic state is found located at 1.35 eV, in good agreement with experiment \cite{wilson}. In the meantime, large linear dichroism (optical birefringence) from anisotropic in-plane crystal structure has been identified. The intrinsic magneto-optical effect from off-diagonal components of dielectric function, however, is found to be very small. CrSBr shows large difference between $\varepsilon_{xx}$ and $\varepsilon_{yy}$ which will dominate the linear dichroism, and consequently the magnitude of rotation angle $\theta$ and ellipticity $\eta$. It has also been revealed that excitonic effect has modified $\varepsilon_{xx}$, $\varepsilon_{yy}$, and $\varepsilon_{xz}$ dramatically, and consequently the linear dichroism and the high-order magneto-optical properties. When the magnetization is tuned by external perpendicular magnetic field towards out-of-plane direction, CrSBr monolayer could exhibit significant Kerr and Faraday effects, which are even larger than those in CrI$_3$ \cite{wu-natcomm} and CrBr$_3$ \cite{wu-prmater}. Furthermore, the short lifetime of the first bright exciton suggests its potential application in infrared light-emitting. Our findings not only have revealed excitonic effect on the optical and magneto-optical properties in 2D ferromagnet CrSBr, but also have shown its potential applications in 2D optics and optoelectronics.

\section{computational method}

Our first-principle calculations are performed using
density-functional theory (DFT) as implemented in the Quantum Espresso package
\cite{qe}. We use the generalized gradient
approximation with Perdew-Burke-Ernzerhof (PBE) \cite{pbe} and norm-conserving pseudopotentials with a plane-wave cutoff
of 80 Ry \cite{nc}. The ground-state wave functions and eigenvalues are calculated within a 
\emph{k} grid of $16\times12\times1$. The structures are relaxed until the total 
forces are less than 0.01 eV/{\AA} and the convergence criterion for 
total energies is set to $10^{-5}$ eV. The quasiparticle band structure and excitonic properties are calculated
with BerkeleyGW package \cite{gw,bse,bgw,bgw-soc} (see Appendix A). We have included spin-orbit coupling with spinor \textit{GW}-BSE calculations \cite{bgw-soc}. A slab model is used
with vacuum layer of 15 {\AA} along the out-of-plane direction and a truncated
Coulomb interaction between CrSBr monolayer and its periodic
image is adopted \cite{sor}. Here, the calculations are based on one-shot $G_0W_0$ with generalized plasmon
pole model. The mean-field wave functions and eigenvalues within PBE 
are chosen as the starting point for $G_0W_0$, 
as a first guess for quasiparticle wave functions and eigenvalues. 
For the convergence of quasiparticle
energies \cite{phzhang}, we have tested the dependence on
\emph{k}-grid size, number of bands, as well as dielectric cutoff.
We use a coarse \emph{k} grid of $16\times12\times1$, 1080 of number of bands, and the dielectric cutoff of
20 Ry (see Appendix B). The electron-hole excitations are then calculated by solving the BSE for each exciton state and frequency-dependent complex dielectric
function $\varepsilon(\omega)$. For the BSE part, the fine
\emph{k} grid of $64\times48\times 1$ is used (see Appendix B). We use a Gaussian smearing with a
broadening constant of 30 meV in optical absorbance spectrum. The
number of bands for optical transitions is 6 valence and 8
conduction bands, which is sufficient to cover the span of the energies of visible light. Such kind of treatment of excited states is robust and has been applied successfully in a wide range of 2D materials \cite{cao1}, including graphene \cite{yang,park,yang1,yang2,yang3,yang4,cao2}, graphyne \cite{huang}, 2D transition metal dichalcogenides \cite{qiu-prl1,yang-apl,qiu-prl2,qiu-prb,yang-prb,jain1,gao,jain2,abr,lu,nai}, black phosphorene \cite{yang-bp,yang-bp-2d,qiu-bp,cao-bp,quek-bp}, blue phosphorene \cite{ju-prapplied-1,ju-prb}, violet (Hittorf's) phosphorene \cite{ju-prresearch,ju-prapplied-2}, 2D monochalcogenides \cite{louie-ges,manos-gese,yanghao}, 2D GaN \cite{manos-gan}, 2D boron nitrides \cite{blase,manos,zhangf}, and 2D magnets \cite{wu-natcomm,wu-prmater,yang5}.

\section{results and discussion}

Bulk CrSBr belongs to an orthorhombic structure with crystal parameters \textit{a} = 3.504 {\AA}, \textit{b} = 4.738 {\AA}, and \textit{c} = 7.907 {\AA} in the space group \textit{Pmmn} (No. 59). This structure is built of monolayers of CrSBr, which are bonded through van der Waals interactions along the \textit{c}-axis. For monolayer CrSBr as demonstrated in Fig. 2, it still belongs to space group \textit{Pmmn} (No. 59) with an effective thickness of around 5.7 {\AA}. The monolayer CrSBr consists of two buckled rectangular planes of CrS fused together, with both surfaces capped by Br atoms. The top view of monolayer CrSBr in Fig. 2 shows a rectangular network lattice with \textit{a} = 3.537 {\AA} and \textit{b} = 4.730 {\AA}. 

\begin{figure}[H]
\centering
\includegraphics[width=0.4\textwidth]{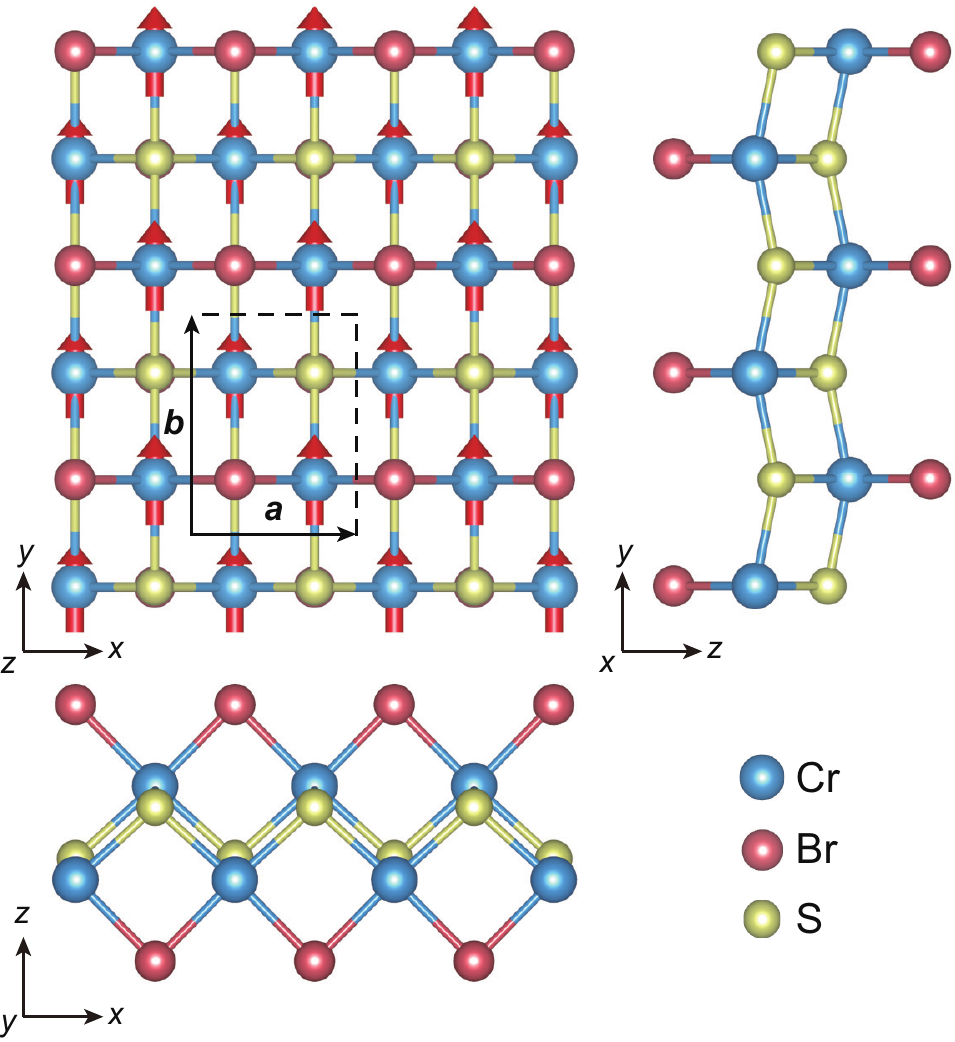}
	\caption{\label{fig:2} Illustration of crystal structure of 2D magnet CrSBr monolayer. Local magnetic moment at Cr atoms pointing along \textit{y} axis.}
\end{figure}

Experimentally, using second-harmonic generation technique, ferromagnetic order with magnetic moment pointing along $y$ axis has been identified in CrSBr monolayer with a Curie temperature of T$_C$ = 146 $\pm$ 2 K \cite{nanoletter-xu}. In the meanwhile, photoluminescence spectra show an obvious decrease at temperature between 130 K and 150 K, suggesting a phase transition therein \cite{wilson}. This Curie temperature in CrSBr monolayer is therefore much higher than Cr$_2$Ge$_2$Te$_6$ monolayer (@ 22 K under magnetic field of 0.075 T) \cite{zhang} and CrI$_3$ monolayer (@ 45 K) \cite{xu}. Theoretically, Cr$^{3+}$ ion has a magnetic moment of 3 $\mu_B$, pointing along the \textit{y} axis of 2D plane. When the magnetization points along \textit{x} or \textit{z} axis, the total energy is 0.042 meV/Cr and 0.072 meV/Cr higher, respectively. And the value along $z$ axis is in agreement with another theoretical result of 0.078 meV/Cr \cite{hwu}. It is noted that the calculated total energy for the state with magnetization along in-plane direction is 0.025 meV/Cr lower in Cr$_2$Ge$_2$Te$_6$ monolayer \cite{guo1} and 0.263 meV/Cr higher in CrI$_3$ monolayer \cite{guo2} when compared with the magnetic state of out-of-plane magnetization. On the other hand, the antiferromagnetic coupling between the in-plane two Cr$^{3+}$ ions is 58.86 meV/Cr higher than the ferromagnetic ground state. This result agrees with previous ground-state calculations \cite{hwu}. For multilayer systems, the local magnetic moment still prefers \textit{y} axis, with magnetic moment antiparallel with each other between neighboring layers \cite{wilson}.

As indicated by the band structure in Fig. 3,
this monolayer of CrSBr is a direct band gap
semiconductor with the top of valence band and the bottom of conduction band both located at $\Gamma$ point. The band gap increases a little along \textit{x} direction and increases sharply along orthogonal direction, i.e., $\Gamma\to\rm{Y}$. Clearly, this monolayer structure shows anisotropic band dispersion. On the other hand, the two spin channels are well separated in the ferromagnetic ground state.

\begin{figure}[H]
\centering
	\includegraphics[width=0.5\textwidth]{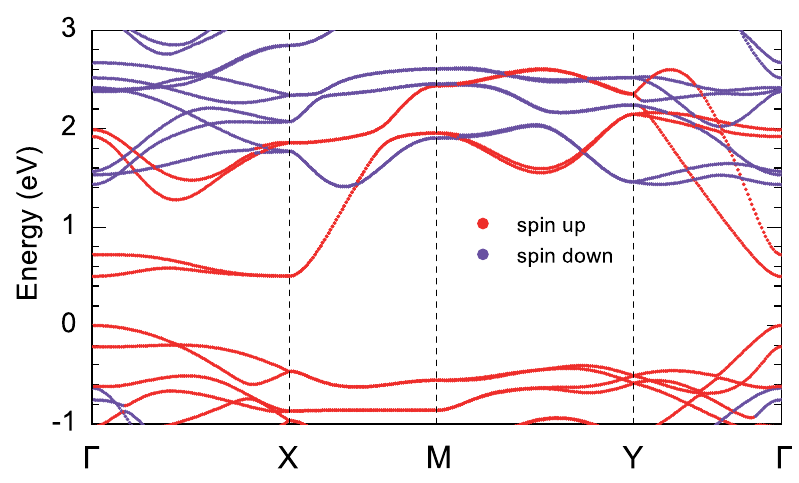}
	\caption{\label{fig:3} PBE band structure with colors denoting the direction of spin polarization along the \textit{y} axis.}
\end{figure}

It should be pointed out that what we have simulated is a suspended monolayer, i.e., a single layer of CrSBr in vacuum. The strength of the Coulomb
interaction in such material originates from weak dielectric
screening in the 2D limit. For distances exceeding few
nanometres, the screening is determined by the immediate
surroundings of the material, which can be vacuum or air in the
ideal case of suspended samples. Compared to MoS$_2$ \cite{qiu-prb} and blue phosphorene \cite{ju-prapplied-1}, such unique 2D dielectric screening in CrSBr should also be anisotropic. The static screened Coulomb interaction is constructed as \cite{qiu-prb}
\begin{equation}
	W_{\rm \bf{GG'}}({\bm{q}}; 0) = \varepsilon^{-1}_{\rm \bf{GG'}}({\bm{q}}; 0) v({\bm{q}}+{\rm \bf{G'}}). 
\end{equation}
The effective static 2D dielectric function $\varepsilon_{2D}(q)$ could be obtained \cite{qiu-prb}
\begin{equation}
	\varepsilon_{2D}^{-1}(\bm{q}) = 
	\frac{|\bm{q}|}{2\pi e^2 L_z}
	\sum_{{\rm \bf G}_z{\rm \bf G'}_z}
	W_{{\rm \bf G}_z{\rm \bf G'}_z}(\bm{q}), 
\end{equation}
where the complicated details of the 
screening in the out-of-plane direction 
$z$ have been integrated out. 
The dielectric screening in such a system obeys particular wavelength dependence. As demonstrated in Fig. 4(a), in the long-distance limit the electron-electron interacts like that in vacuum with $\varepsilon=1$, while within the intermediate distance, the electron-electron interaction has effective dielectric screening by the 2D materials with $\varepsilon \ge 1$. The label ``others" in Fig. 4(a) refers to the other $\bm{q}$ points in $q$ mesh which do not belong to ($q_x$,0) or (0,$q_y$). For the anisotropic nature, the dielectric screening changes along different directions. To accurately access the distance-dependence of 2D dielectric screening, we have obtained $\varepsilon_{\rm 2D}(\bm{s})$ through Hankel transformation of $\varepsilon_{\rm 2D}(\bm{q})$ and shown it in Fig. 4(b). Clearly, if two charges are very close together, there is not enough space for the electronic cloud to polarize, so
$\varepsilon_{\rm 2D}(\bm{s}\rightarrow0)$ $=$ $1$. On the other hand, if the two charges are very far away, the field lines connecting the charges travel mainly
through the vacuum, so they are not much affected by the
intrinsic dielectric environment of the quasi-2D semiconductor
and  $\varepsilon_{\rm 2D}(\bm{s}\rightarrow\infty)$ $=$ $1$. Between these two ends, $\varepsilon_{\rm 2D}(\bm{s})$ is influenced by the intrinsic material property of CrSBr monolayer and is larger than 1. At finite distance $\bm{s}_{max}$, $\varepsilon_{\rm 2D}(\bm{s}_{max})$ will exhibit its maximum. Such kind of 2D dielectric screening has also been revealed in MoS$_2$ monolayer with $\bm{s}_{max} = 1.5$ $\rm \AA$ and $\varepsilon_{\rm 2D}(\bm{s}_{max})$ = 11 \cite{qiu-prb}. For CrSBr monolayer, $\bm{s}_{max}$ is little larger and $\varepsilon_{\rm 2D}(\bm{s}_{max})$ is weaker. In the meantime, due to anisotropic crystal structure, the 2D dielectric screening in CrSBr monolayer also exhibits strong anisotropy. $\varepsilon_{\rm 2D}(\bm{s}_{max})$ is 4.4 and 7.0 for $\varepsilon_{\rm 2D}(\bm{s})$ along $x$ and $y$ axis, respectively. And corresponding $\bm{s}_{max}$ is 3.0 $\rm \AA$ and 4.0 $\rm \AA$, the same order with lattice constants. Interestingly, such kind of anisotropy persists to long distance. As shown in the inset of Fig. 4(b), the difference between $\varepsilon_{\rm 2D}(\bm{s})$ along $x$ and $y$ axis does not vanish until 60 $\rm \AA$. So at the distance from 1 $\rm \AA$ to ten times length of lattice constant, the anisotropic
dielectric screening dominates Coulomb interaction in CrSBr monolayer. For 2D system, Coulomb potential could be described by the Keldysh model \cite{kel}, where the potential between two charges has the form
\begin{equation}
	V_{\rm 2D}(\bm{s}) = \frac{\pi e^2}{2\rho_0}[H_0(\frac{|\bm{s}|}{\rho_0})-Y_0(\frac{|\bm{s}|}{\rho_0})].  
\end{equation} 
Here $H_0$ and $Y_0$ are, respectively, the Struve and Bessel
functions of the second kind. $\rho_0$ is a screening length, which
is $\rho_0 = 2 \pi \alpha_{\rm 2D}$, and $\alpha_{\rm 2D}$ is the 2D polarizability. Taking the 2D Fourier
transform of Eq. 5 results in a dielectric function of the form
\begin{equation}
	\varepsilon_{\rm 2D}(\bm{q}) = 1 + \rho_0 |\bm{q}|.
\end{equation}
We fit the Keldysh model to our $ab$ $initio$ effective
dielectric function at small $\bm{q}$, where the Coulomb potential approaches 1/s. The 2D polarizability, however, shows large difference between orthogonal two directions. $\alpha_{\rm 2D}$ is 4.2 $\rm \AA$ for $\bm{q}$ along $x$ axis and 10.4 $\rm \AA$ for $\bm{q}$ along $y$ axis. Based on above discussion, in sufficiently long range from 1 $\rm \AA$ to 60 $\rm \AA$, the dielectric screening in CrSBr monolayer is anisotropic.

\begin{figure}[H]
\centering
	\includegraphics[width=0.5\textwidth]{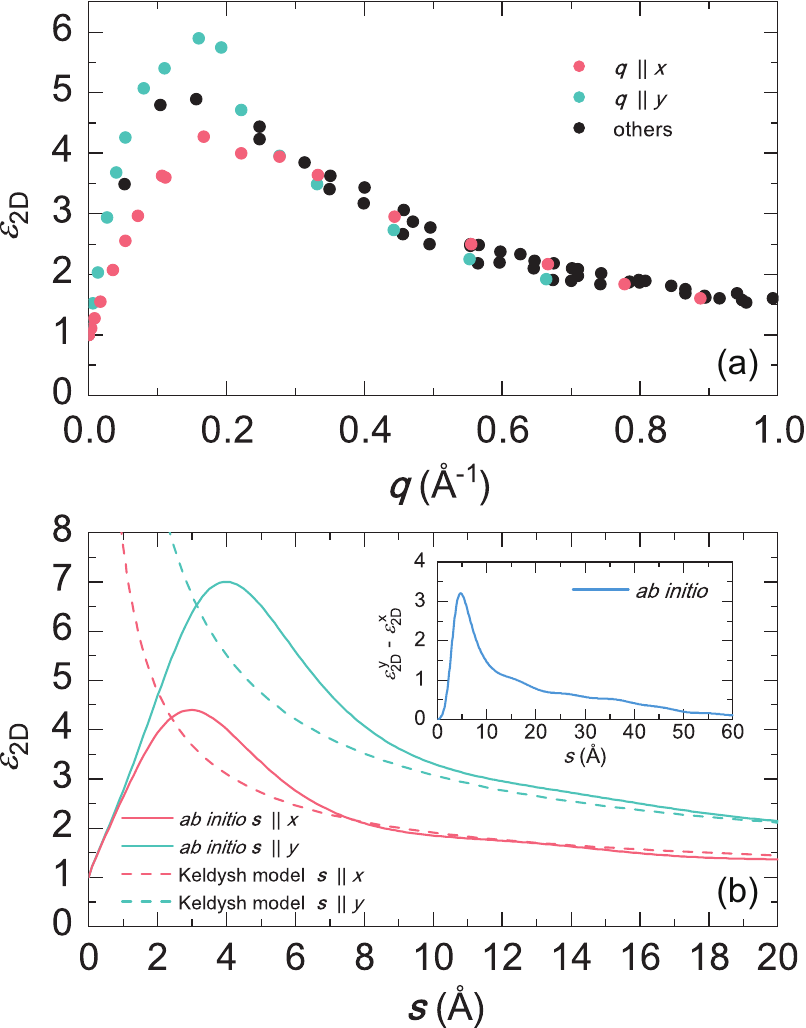}
	\caption{\label{fig:4} (a) Frequency-dependence of 2D dielectric
		screening in CrSBr monolayer under $ab$ $initio$ calculation. (b) Distance-dependence of 2D dielectric
		screening in CrSBr monolayer under $ab$ $initio$ calculation (solid line) and Keldysh model (dashed line). Inset of (b) shows the difference between $\varepsilon_{\rm 2D}(\bm{s})$ along $y$ axis and $x$ axis.}
\end{figure}

Due to the unique dielectric environment (monolayer 2D material suspended in vacuum or other dielectric surroundings), electron-electron and electron-hole interactions in 2D materials are much stronger than conventional bulk materials, like GaAs, Si, and so on. Therefore, dielectric screening in 2D materials is reduced compared with conventional bulk materials. With the reduced dielectric screening in 2D CrSBr monolayer, the quasiparticle correction to the electronic band structure is large. At $\Gamma$ point, the quasiparticle band gap is 2.22 eV, where the value is 0.50 eV within PBE. In Fig. 5, we have shown the direct quasiparticle band gap in the Brillouin zone. The one-dimensional nature of the lowest transition energies is obvious. 

\begin{figure}[H]
\centering
	\includegraphics[width=0.4\textwidth]{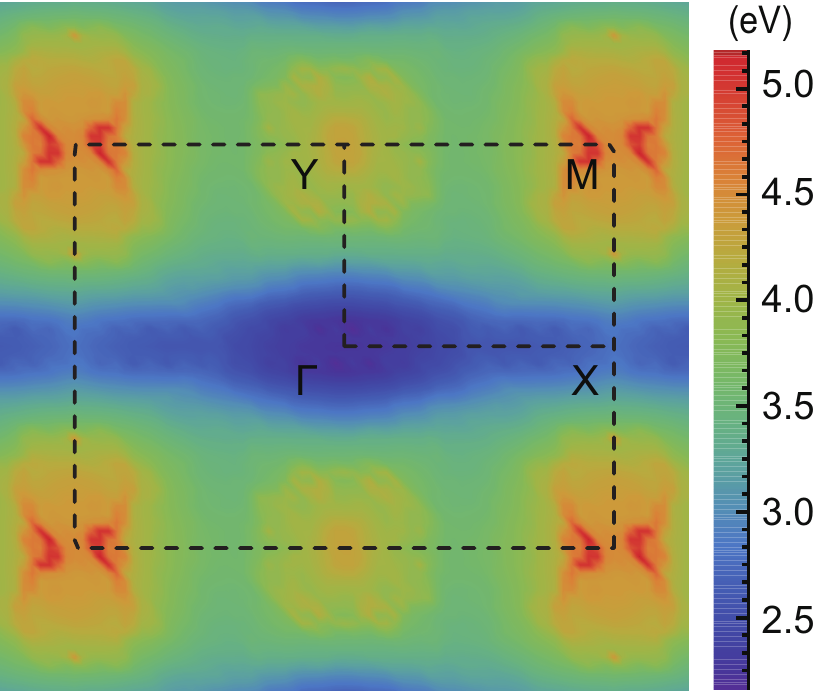}
	\caption{\label{fig:5} The lowest valence to conduction band transition energy (based on quasiparticle energies).}
\end{figure}

The optical absorbance in CrSBr monolayer is further calculated and shown in Fig. 6, where the difference between two in-plane orthogonal directions is obvious. Although the main optical absorption is located at 3.01 eV, the small peak located below could catch as high as 20 \% of the incident light. The optical absorption edge is located at 1.35 eV when the electric field is along \textit{y} direction. For the other direction, the absorption edge is 1.73 eV. Compared with the results without the consideration of electron-hole interaction, the excitonic effect is obvious in this 2D material. In Fig. 6(c), the exciton states are demonstrated and agree with the optical spectrum well. Dark excitons are also indicated in these non-Rydberg series. Clearly, the binding energy for the first exciton state could reach as high as 1.0 eV, almost 1/3 of the quasiparticle band gap for the center of electron-hole excitations. When the electric field is rotated along the \textit{xy} plane, as shown in Fig. 7 for several bright exciton states, the oscillator strength shows an anisotropic distribution.

\begin{figure}
	\includegraphics[width=0.55\textwidth]{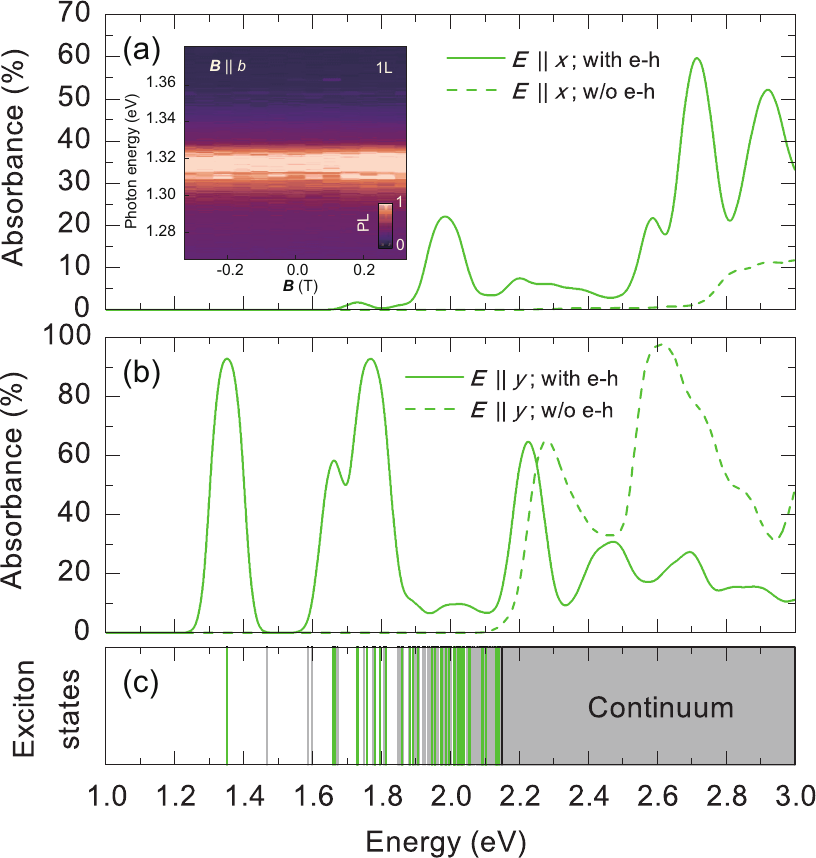}
	\caption{\label{fig:6} Anisotropic optical absorbance with electric polarization along \textit{x} axis (a) and \textit{y} axis (b) and corresponding exciton spectrum (c) in CrSBr monolayer. Here, green lines are for the bright excitons and the gray lines are for dark excitons. The dotted lines are for the optical absorption based on IPA. Inset shows the experiment data of PL spectrum of CrSBr monolayer under magnetic field along \textit{y} axis \cite{wilson}.}
\end{figure}

\begin{figure}
	\includegraphics[width=0.6\textwidth]{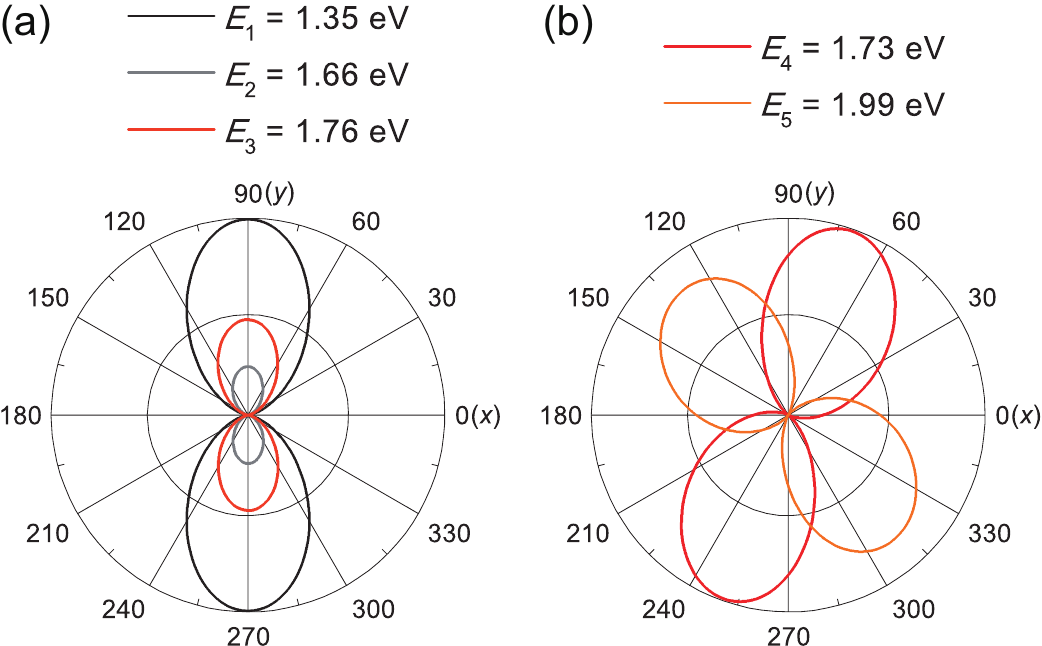}
	\caption{\label{fig:7} Optical anisotropy (absorbance amplitude) for the five excitonic states with the wavelength of 1.35 eV (918 nm), 1.66 eV (747 nm), 1.76 eV (705 nm), 1.73 eV (716 nm), and 1.99 eV (623 nm), respectively.}
\end{figure}

\begin{figure}
	\includegraphics[width=0.9\textwidth]{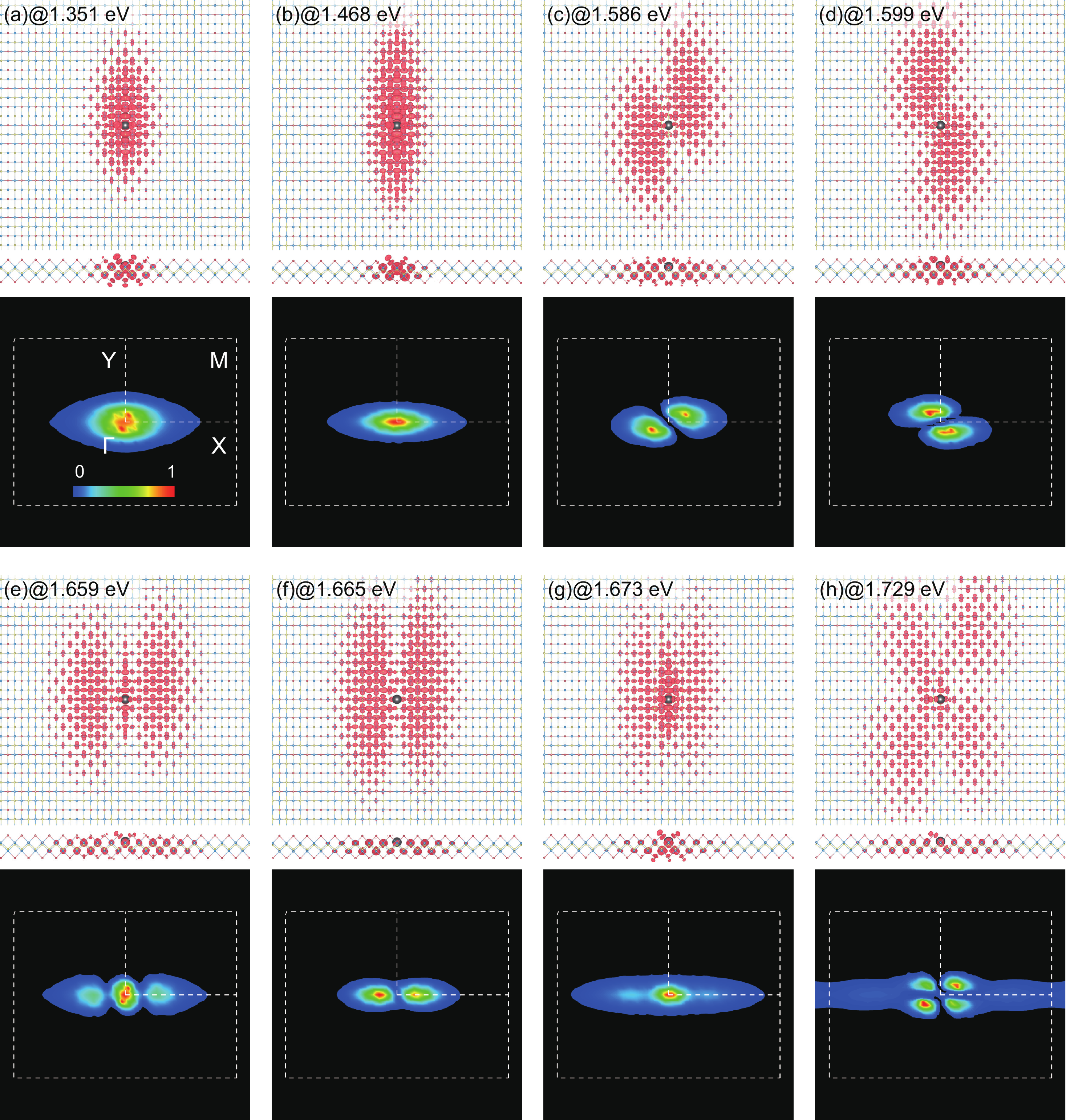}
	\caption{\label{fig:8} Real-space plots (upper panel) of modulus squared of
		the exciton wave function and corresponding reciprocal-space plots (down panel) for the first eight excitons. Here, the hole (black circle) is fixed at Cr.}
\end{figure}

We now look at the character of the first eight excitons, which are all excited around the $\Gamma$ point. In Fig. 8, we show the modulus squared of the real-space exciton wave function when the hole is fixed near a Cr atom. Electron-hole pair amplitudes (the envelope functions) of low energy exciton wave functions in reciprocal space are plotted in the down panel of each figure. From these plots, the nodal structures of the envelope functions of the states are apparent. $1s$ state has only one node with the strongest transition at $\Gamma$ point. In detail, in Fig. 8(a), the character of the exciton corresponding to the first peak in the absorption spectrum (E$\parallel$\textit{y}) reflects the transitions from the top of valence band to the lowest conduction band around $\Gamma$ point. The envelope of the exciton wave function is almost azimuthally symmetric. The root-mean-square radius of the exciton in real space is around 1.4 nm. The second excitonic state is a dark exciton. It is also a 1\textit{s} state and also arises from the electron-hole excitations around $\Gamma$ point, but the orbitals for the electron-hole formation are different from the first bright state. Although it has a much smaller oscillator strength, the electric polarization along \textit{y} direction is much higher than that along the \textit{x} direction. The next two dark excitons shown in Fig. 8(c) and (d) are $2p$ states. Bright exciton shown in Fig. 8(e) is 2\textit{s} state and the exciton in Fig. 8(h) is 3\textit{d} state. Together with the first state, these five states could be categorized into the series of electron-hole excitations from v1 to c1 mainly at $\Gamma$ point. This can also be judged from energy distribution in Fig. 9, where the distribution of the constituent free electron-hole pairs specified by ($E_v$, $E_c$) for selected exciton states weighted by the module squared exciton envelop function for each specific interband transition is plotted for all these eight excitonic states. Clearly, these five states have similar patterns. As shown in Appendix C, the orbitals corresponding to the electron-hole pairs are Cr $d_{yz}$ and Cr $d_{x^2-y^2}$ \cite{w90}. For the second and sixth excitons, it involves v1 to c2 transitions, i.e., from Cr $d_{yz}$ to Cr $d_{x^2-y^2}$ hybrid with S $3p$. These transitions bring to the above mentioned dark 1\textit{s} state in Fig. 8(b) and the bright 2\textit{p} state in Fig. 8(f). For the exciton shown in Fig. 8(g), it involves more complicated electron-hole pairs formation (from v2 to c1 and v1 to c2 as indicated in Fig. 9(g)), and it is a dark state. Characterization of all the node structures of these excitons is summarized in Table I. Obviously, these excitons process an increased real-space extension of the wave function. The effective electron-hole interaction is therefore reduced strongly, consistent with the smaller binding energy as shown in Table I. The difference in the oscillator strength between \textit{x} direction and \textit{y} direction is obvious for all the exciton states. 

\begin{figure}
	\includegraphics[width=0.9\textwidth]{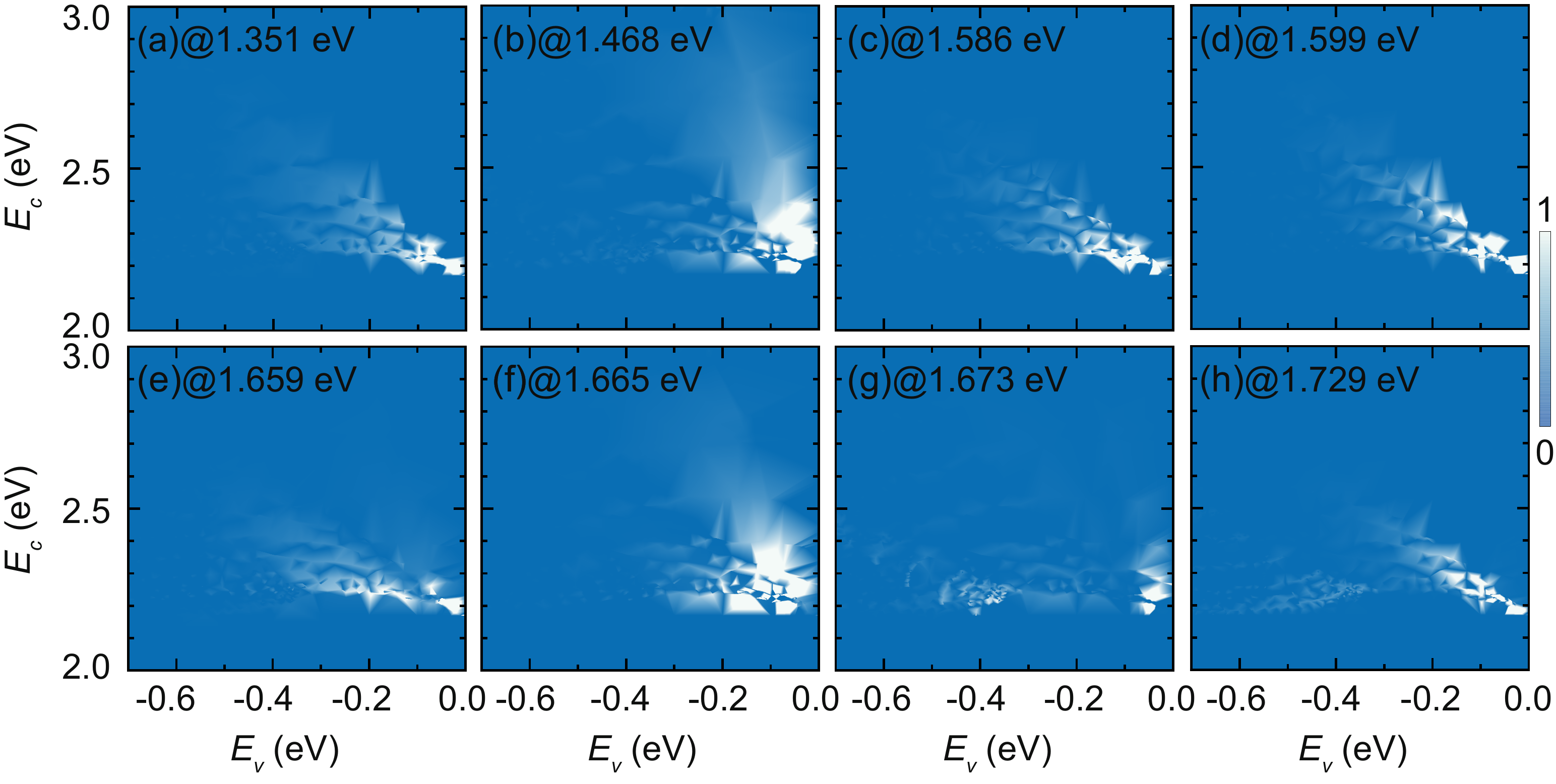}
	\caption{\label{fig:9} The distribution of free electron-hole pair with electron energy at $E_c$ and hole energy at $E_v$ for selected exciton states weighted by module squared exciton envelope function for each interband transitions between states $|\textbf{vk}\rangle$ and $|\textbf{ck}\rangle$, with quasiparticle energies of $E^{\rm QP}_{vk}$ and $E^{\rm QP}_{ck}$, respectively.}
\end{figure}

\begin{table}[H]
\centering
\caption{Excitonic properties of first eight excitons with excitation energy ($E_{xct}$), exciton binding energy ($E_b$), and oscillator strength for electric field along $x$ and $y$ directions. }
\begin{ruledtabular}
\begin{tabular}{cccll}
		\multirow{2}{*}{Excitons} &     \multirow{2}{*}{$E_{xct}$ (eV)}      &    \multirow{2}{*}{$E_b$}  & \multicolumn{2}{c}{Oscillator strength} \\
		& &  & \;  \textbf{\textit{E}} $\parallel$ \textit{x} & \; \textbf{\textit{E}} $\parallel$ \textit{y}  \\
		\hline
		1 &   1.351  &  0.870 &  $1.9 \times10^{-3}$  &  $2.3 \times10^{4}$  \\
		2 &   1.468  &  0.735 &   $2.7 \times10^{-6}$ &  $3.2 \times10^{-4}$ \\
		3 &   1.586  &  0.635&   $3.6 \times10^{-2}$  & $4.7 \times10^{0}$ \\
		4  &   1.599  &  0.622&   $2.8 \times10^{-3}$  & $4.1 \times10^{0}$ \\
		5 &    1.659  &  0.562 &  $6.4 \times10^{-1}$  & $5.8 \times10^{3}$ \\
		6 &   1.665  &  0.556 &   $1.2 \times10^{-1}$  & $4.6 \times10^{2}$  \\
		7 &  1.673  &  0.548 &  $3.8 \times10^{-4}$  & $5.0 \times10^{-1}$ \\
		8 &   1.729  &  0.492 &  $8.3 \times10^{1}$  & $3.8 \times10^{2}$  \\
\end{tabular}
\end{ruledtabular}
\end{table}

With the above revealed anisotropic quasiparticle electronic band structure, electron-hole excitation, and optical absorption, we continue to show the (magnetic) linear dichroism and MO SH and Vogit effects in CrSBr monolayer. The essence of a theoretical modeling of the MO effects lies in accurately accounting for the diagonal and off-diagonal frequency-dependent macroscopic dielectric functions, which have been readily available from our \textit{$G_0W_0$}-BSE calculations with electron-hole interaction included. As verified by the agreement between our theoretical calculations and experiments of CrSBr as well as the fact that exciton effect in ferromagnetic monolayer CrI$_3$ and CrBr$_3$ has strongly modified its MO responses \cite{wu-natcomm,wu-prmater}, significantly different behaviors going beyond those from a treatment considering only transitions between non-interacting Kohn-Sham orbitals should be expected in CrSBr. To simulate the experimental setup, as demonstrated in Fig. 1, we consider CrSBr monolayer deposited on a dielectric substrate $\alpha$-Al$_2$O$_3$, which has a wide band gap of 8.7 eV with refraction constant of 1.75 \cite{wu-prmater}. Assuming a linearly polarized incident light, we calculate the SH and Voigt signals by analyzing the polarization plane of the reflection (transmission) light, which is in general elliptically polarized with a rotation angle and an ellipticity. The detailed calculations are based on the transfer matrix method and could be found in Appendix D. In Fig. 10, we have found that the rotation angle could be larger than 10$^{\circ}$ for the reflected light. For transmitted light, the largest rotation angle is around 8$^{\circ}$. On the other hand, the revealed SH and Voigt effects dominate within the quasiparticle band gap. When the electron-hole interaction is not considered, both the amplitude and the position of the spectrum are modified significantly (see Fig. 15 in Appendix E). This verifies again the fact that it is the exciton effect that leads to the large linear dichroism in CrSBr monolayer. In addition, we can find that the off-diagonal component of dielectric function plays little role in the above revealed large linear dichroism in CrSBr, where as shown in Fig. 16 (Appendix E), $\theta_{\rm SH}$ and $\theta_{\rm V}$ do not change. 

\begin{figure}[H]
	\centering
	\includegraphics[width=0.5\textwidth]{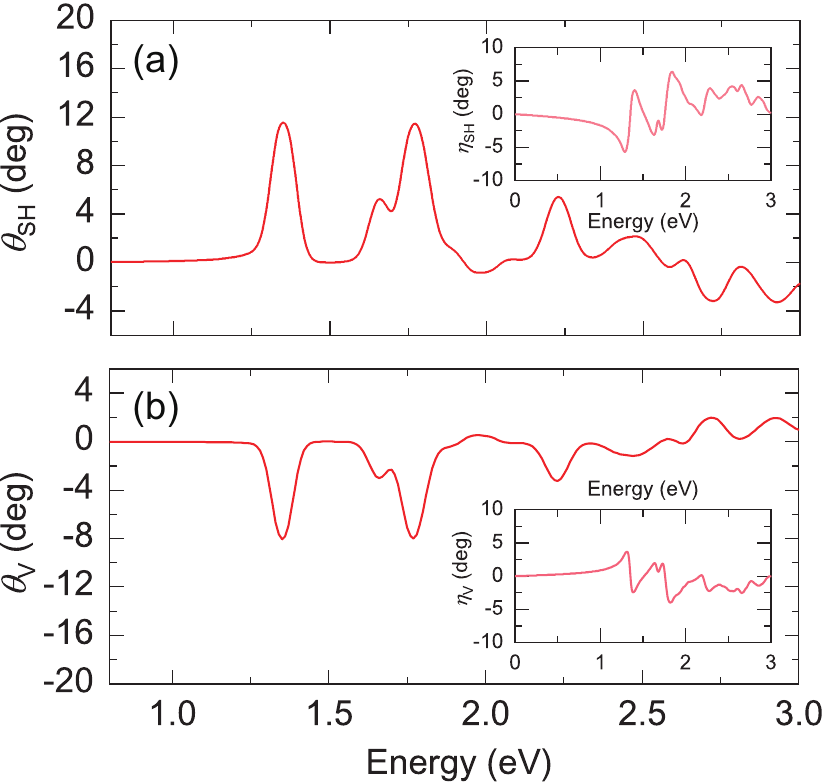}
	\caption{\label{fig:10} Rotation angles of (a) reflected and (b) transmitted light in CrSBr monolayer. Insets are ellipticity.}
\end{figure}

We have also performed the calculations of magneto-optical effects when spin orientation is pointing along $x$ and $z$ axis. As shown in Fig. 17 and Fig. 18 (Appendix F), absorbance spectrum and exciton states are similar to those for the state with spin orientation along $y$ axis. In the meantime, due to extremely small $\varepsilon_{yz}$, the revealed rotation angle of long axis of the polarization ellipse for reflection and transmission light is still determined by the large ratio of $\varepsilon_{xx}$/$\varepsilon_{yy}$ (see Fig. 19 in Appendix G). On the other hand, when spin orientation is tuned by external perpendicular magnetic field towards out-of-plane direction, i.e. $z$ axis, the magneto-optical Kerr and Faraday effects will appear in CrSBr monolayer. In this case, a linearly polarized continuous-wave
light is modified by the presence of out-of-plane magnetic field when propagating through CrSBr monolayer. The left and right-circularly polarized components will propagate with different refractive index and will pick up different optical path length and absorption. Therefore, the
reflected or transmitted light becomes elliptically polarized
(characterized by an ellipticity $\eta$), and the long axis of
the polarization ellipse is rotated (characterized by a rotation
angle $\theta$). The detailed theoretical calculations of rotation angles and ellipticities are shown in Appendix H. Meanwhile, as shown in Fig. 21 (Appendix H), the maximal rotation angles of Kerr and Faraday effects are 1.1$^\circ$ and 0.6$^\circ$, respectively, which are even larger than 0.9$^\circ$ and 0.3$^\circ$ in CrI$_3$ \cite{wu-natcomm} and 0.3$^\circ$ and 0.2$^\circ$ in CrBr$_3$ \cite{wu-prmater}. Similar to the cases with in-plane magnetization, the optical spectrum has been modified significantly for the exciton effects (see Fig. 22 in Appendix H). We list the maximal rotation angles in TABLE \uppercase\expandafter{\romannumeral2} for different spin orientations. Therefore, for both linearly and circularly polarized light, the large phase shift facilitates the applications of CrSBr monolayer as optical polarizers and waveplates.

\begin{table}[H]
	\centering
	\caption{Comparison of magneto-optical effects for reflection and transmission lights, polarization of lights propagating in CrSBr monolayer, off-diagonal elements, and maximums of corresponding rotation angles with electron-hole interaction when spin orientation is along $x$, $y$, and $z$ axis in the energy range of 0 $\sim$ 3 eV. }
	\begin{ruledtabular}
		\begin{tabular}{ccccc}
			{Spin orientation}   &    {Magneto-optical effects} &  Polarization  &    {Off-diagonal elements}      & {Rotation angles ($^\circ$)}   \\
			\hline 
			$x$  &  SH/Voigt   & linear & $\varepsilon_{yz}$  &  11.6/8.1  \\
			$y$  &  SH/Voigt  & linear & $\varepsilon_{xz}$  &  11.6/8.1 \\
			$z$  &  Kerr/Faraday & circular  &  $\varepsilon_{xy}$  &  1.1/0.6  \\
		\end{tabular}
	\end{ruledtabular}
\end{table}

For black phosphorene, it is noted that the ratio between the long axis and short one is 1.4 for the anisotropic crystal structure and 1.5 eV energy difference between absorption edges for the electric field along two axes could be found (see Fig. 23 in Appendix I). For comparison, it is noted that the rotational angle in black phosphorene is of similar magnitude. However, CrSBr shows two broad peaks at the energies of both infrared and red lights, whereas black phosphorene exhibits a single narrow peak in this energy window (see Fig. 24 in Appendix I). Additionally, obvious linear dichroism has been revealed experimentally in black phosphorene and its few-layers \cite{sch,xia2,cui,fwang}. Therefore, polarization-sensitive broadband photodetector using a CrSBr vertical \textit{p-n} junction could be constructed successfully for the application in 2D semiconductor optoelectronics. 

To demonstrate other potential applications in 2D semiconductor optoelectronics, we further calculate the lifetime of the first bright exciton.  Using ``Fermi's golden rule", the radiative lifetime $\tau_{S}(0)$ at
0 K of an exciton in state $S$ is derived according to \cite{lifetime,palummo-nanolett}
\begin{equation}
	\tau_S(0)=\frac{\hbar^{2}c}{8\pi e^{2}E_{S}(0)}\frac{A_{\rm	{uc}}}{\mu_{S}^2},
\end{equation}
where $c$ is the speed of light, $A_{\rm {uc}}$ is the area of the
unit cell, $E_{S}(0)$ is the energy of the exciton in state
\emph{S}, and  $\mu_{S}^2=(\hbar^2/m^{2}E_{S}(0)^2) 
(|\left\langle  G|p_{||}|\Psi_{S}\right\rangle|^{2}/N_k)$ is the square modulus of the BSE exciton transition
dipole divided by the number of unit cells in this 2D system. The
exciton radiative lifetime $\left\langle\tau_{S}\right\rangle$
at temperature \emph{T} is obtained as
\begin{equation}	
	\left\langle\tau_{S}\right\rangle=\tau_{		
		S}(0)\frac{3}{4}(\frac{E_{S}(0)^2}{2M_{S}c^2})^{-1}k_{\rm	
		B}T,
\end{equation}
where $k_{\rm {B}}$ is Boltzmann constant, and $M_{S}=m_e^*+m_h^*$ is the exciton effective mass. The computed radiative lifetime of the first bright exciton in CrSBr monolayer at 4 and 300 K is 0.001 ns and 0.100 ns, respectively, which are even smaller than those in the conventional transition metal dichalcogenides, e.g., 0.004 ns and 0.270 ns in MoS$_2$ \cite{palummo-nanolett}. In Table \uppercase\expandafter{\romannumeral3}, we have listed the exciton lifetime of some typical 2D semiconductors \cite{quek-bp,ju-prapplied-1,ju-prresearch,manos-gan,palummo-nanolett,Yang-Light,cui-Small,korn-apl,Lagarde-prl,Shi-acsnano,Robert-prb,Kumar-prb,Wang-prb,Mouri-prb}, as well as conventional semiconductor GaInN/GaN quantum wells \cite{Langer-apl}. Clearly, such a short lifetime in CrSBr monolayer together with its direct band gap shows its advantages in infrared light-emitting applications. Moreover, as shown in Fig. 11, the strong optical absorption of CrSBr monolayer covering the whole solar spectrum as well as its optimized band gap (1.35 eV) within the Shockley-Queisser limit also suggests efficient solar energy conversion. Therefore, from gapless graphene to narrow-band-gap black phosphorene, intermediate-band-gap transition-metal dichalcogenides, and wide-band-gap semiconductors of Hittorf’s phosphorene, blue phosphorene, \uppercase\expandafter{\romannumeral3}-\uppercase\expandafter{\romannumeral5} monochalcogenides, and boron nitride, magnetic CrSBr could play an important role in infrared optoelectronics and 2D photovoltaics.

\begin{table}[H]
	\centering
	\caption{Collection of exciton radiative lifetime $\tau$ (ns) for various 2D materials. $\tau_{\rm S}^{\rm LT}$ and $\tau_{\rm S}^{\rm RT}$ are the computed radiative lifetimes at low temperature ($\approx$ 4 K) and room temperature, respectively. Experimental data of $\tau_{\rm exp}^{\rm LT}$ and $\tau_{\rm exp}^{\rm RT}$ are also listed for comparison. }
	\begin{ruledtabular}
		\begin{tabular}{ccccc}
			& {$\tau_{\rm S}^{\rm LT}$}   &    {$\tau_{\rm S}^{\rm RT}$}   &    {$\tau_{\rm exp}^{\rm LT}$}      & {$\tau_{\rm exp}^{\rm RT}$}   \\
			\hline 
			CrSBr monolayer  &  0.001   &  0.100  &   &  \\
			Black phosphorus monolayer  &  $3\times10^{-5}$ \cite{quek-bp}  &  0.002 \cite{quek-bp}  &   & 0.221 \cite{Yang-Light} \\
			Hittorf’s phosphorus monolayer &  0.24 \cite{ju-prresearch}  &  18 \cite{ju-prresearch}  &   &  \\
			GeS monolayer &  $7\times10^{-5}$ \cite{quek-bp} &  0.005 \cite{quek-bp}  &   &  \\
			ReS$_2$ monolayer &   &    &   & 0.04 \cite{cui-Small} \\
			Blue phosphorus monolayer &  0.003 \cite{ju-prapplied-1}  &  0.24 \cite{ju-prapplied-1}  &  &  \\
			MoS$_{2}$ monolayer &  0.004 \cite{palummo-nanolett}  &  0.27 \cite{palummo-nanolett}  & 0.005 \cite{korn-apl}, 0.005 \cite{Lagarde-prl}  & 0.85 \cite{Shi-acsnano} \\
			MoSe$_{2}$ monolayer &  0.005 \cite{palummo-nanolett}  &  0.38 \cite{palummo-nanolett}  & 0.002 \cite{Robert-prb}  & 0.9 \cite{Kumar-prb} \\
			WS$_{2}$ monolayer &  0.002 \cite{palummo-nanolett}  &  0.17 \cite{palummo-nanolett}  &   &  \\
			WSe$_{2}$ monolayer &  0.004 \cite{palummo-nanolett}  &  0.29 \cite{palummo-nanolett}  & 0.002 \cite{Robert-prb}, 0.004 \cite{Wang-prb}  & 4 \cite{Mouri-prb} \\
			2D GaN &    &    & 0.6 \cite{manos-gan}  &  \\
			GaInN/GaN QW &    &    &  & $10^{2}$ \cite{Langer-apl} \\
		\end{tabular}
	\end{ruledtabular}
\end{table}

\begin{figure}[H]
	\centering
	\includegraphics[width=0.7\textwidth]{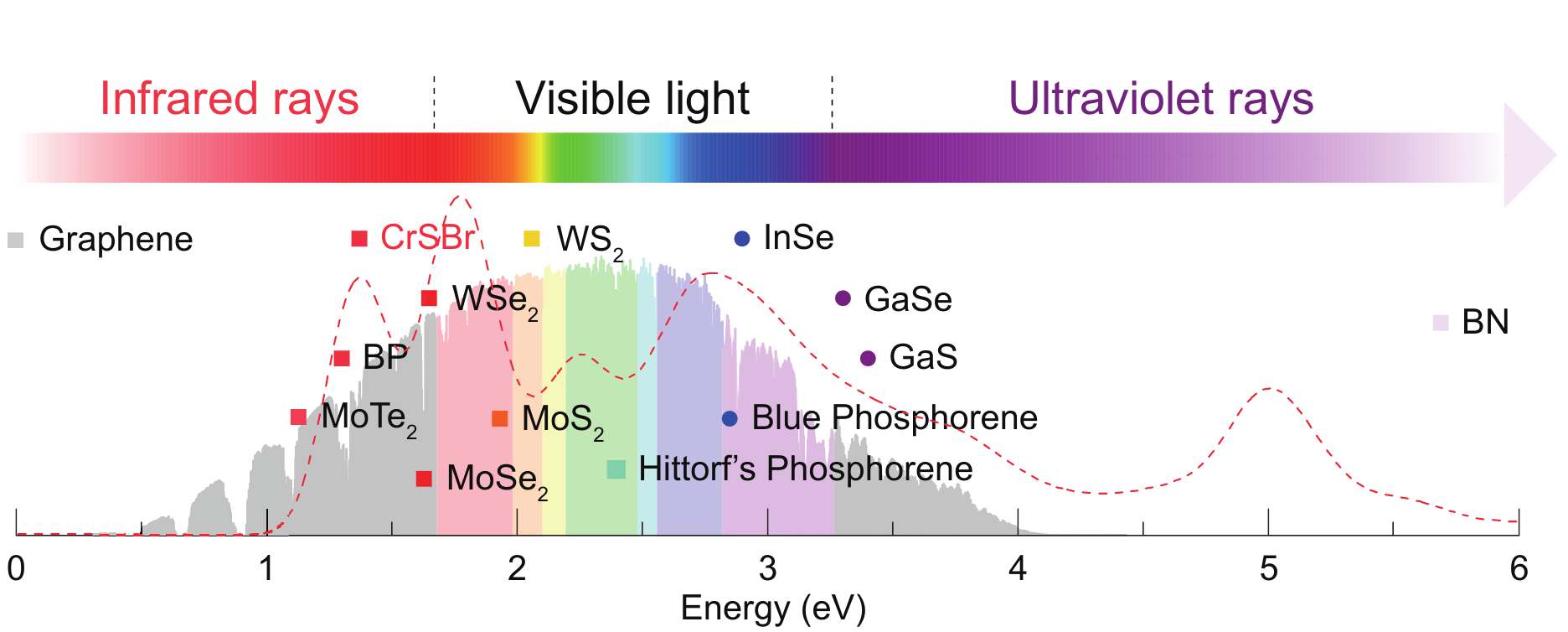}
	\caption{\label{fig:11} Some 2D materials covering a broad spectral range,
		from gapless graphene to narrow-band-gap black phosphorene, intermediate-band-gap transition-metal
		dichalcogenides, and wide-band-gap semiconductors of Hittorf’s phosphorene, blue
		phosphorene, \uppercase\expandafter{\romannumeral3}-\uppercase\expandafter{\romannumeral5} monochalcogenides, and boron nitride. Squares and circles represent direct and indirect band gap semiconductors, respectively. The
		solar spectrum \cite{solar-spectrum} and averaged absorbance (dashed red line) of light polarized along $x$  and $y$ axis in CrSBr monolayer are also shown.}
\end{figure}

\section{summary}

To summarize, by considering explicitly many-body effects of the electron-electron and electron-hole interactions in 2D ferromagnet CrSBr, we have demonstrated unusual large optical anisotropy in this 2D material. The inherent in-plane magnetization with the consideration of spin-orbit coupling is found to contribute little to the revealed large linear dichroism. The giant SH and Voigt effects, which are comparable with those in black phosphorene, originate from the inherent anisotropic crystal structure and coherent one-dimensional band dispersion for anisotropic electron-hole pairs excitation. Our calculation of the first exciton located at 1.35 eV agrees with a recent experiment of photoluminescence. The in-gap exciton states of either bright or dark show the diversity of electron-hole excitation in this 2D magnet. Compared with black phosphorene, the relative delocalization of exciton in \textit{k}-space indicates the nature of electron-hole excitation in CrSBr is in-between Wannier-Mott type and Frenkel type. Excitonic effect has modified $\varepsilon_{xx}$, $\varepsilon_{yy}$, and $\varepsilon_{xz}$ dramatically, and consequently the linear dichroism and the magneto-optical properties. With out-of-plane magnetization, Kerr and Faraday effects have been revealed in CrSBr monolayer, which are even larger than those in CrI$_3$ and CrBr$_3$. Furthermore, the short exciton lifetime as well as the strong optical absorption covering the whole solar spectrum shows its advantages in application of light-emitting diode and solar cell. Our studies have provided a basic framework to account for the high-order magneto-optical effects in 2D magnets and also shown potential applications of CrSBr in 2D optics and optoelectronics.

\

\begin{acknowledgments} S. J. thanks Prof. Steven G. Louie and his group for their great help on BerkeleyGW method and useful discussion of exciton physics in 2D materials. This work is supported by the National Basic Research
	Program of China (973 Program 2019YFA0308402) and the National Natural Science Foundation
	of China under Grant No. 51972217. T. X. Q. and J. Z. contributed equally to this work.
\end{acknowledgments}

\appendix

\section{BASIC FRAMEWORK OF \textit{GW}-BSE IN BERKELEYGW PACKAGE}

Generally, the quasiparticle self-energies 
are obtained by solving the Dyson equation \cite{gw}. \begin{equation}
	\left[ -\frac{1}{2}\nabla^2 + V_{\rm ion} + V_{\rm H} +
	\Sigma\left( E^{\rm QP}_{n{\rm \bf{k}}} \right) \right]\psi^{\rm
		QP}_{n{\rm \bf{k}}} = E^{\rm QP}_{n{\rm \bf{k}}} \psi^{QP}_{n{\rm
			\bf{k}}},
\end{equation} where $\Sigma$ is the self-energy operator within the \emph{GW}
approximation, and $E^{QP}_{n{\rm \bf{k}}}$ and $\psi^{QP}_{n{\rm
		\bf{k}}}$ are the quasiparticle energies and wavefunctions,
respectively. The self-energy operator $\Sigma$ is invoked for the quasiparticle behavior with quasiparticle energies $E^{\rm QP}_{n{\rm \bf{k}}}$ and wave functions $\psi^{\rm QP}_{n{\rm\bf{k}}}$. Here, we use the many-body perturbation method with one-shot $G_0W_0$ framework. The mean-field wave functions within DFT-PBE are used as quasiparticle wave functions and the quasiparticle energy is approached starting from DFT-PBE eigenvalue. 

We first compute static polarizability based on random-phase
approximation (RPA) \cite{gw}: 
\begin{equation}
	\chi_{\rm \bf{GG'}}({\rm \bf{q}}; 0) = \sum_{n, n', \rm \bf{k}}
	\langle n,{\rm \bf{k}} | e^{-i({\rm \bf{q}}+{\rm \bf{G}})\cdot{\rm
			\bf{r}}} | n',{\rm \bf{k}} + {\rm \bf{q}} \rangle \langle n', {\rm
		\bf{k}} + {\rm \bf{q}} | e^{i({\rm \bf{q}}+{\rm \bf{G'}})\cdot{\rm
			\bf{r}}'} | n, {\rm \bf{k}} \rangle \frac{1}{E_{n'{\rm \bf{k}}+{\rm
				\bf{q}}}-E_{n{\rm \bf{k}}}},
\end{equation}
where $n$, $n'$ are occupied and unoccupied band indices, $\rm
\bf{k}$ is wave vector, $\rm \bf{q}$ is a vector in the first
Brillouin zone, $\rm \bf{G}$ is a reciprocal-lattice vector, and
$|n, \rm \bf{k}\rangle$ and $E_{n\rm \bf{k}}$ are the mean-field
electronic eigenvectors and eigenvalues, respectively.

Then the dielectric matrix is constructed as \cite{gw}
\begin{equation}
	\varepsilon_{\rm \bf{GG'}}({\rm \bf{q}}; 0) = \delta_{\rm \bf{GG'}} - v({\rm
		\bf{q}}+{\rm \bf{G}}) \chi_{\rm \bf{GG'}}({\rm \bf{q}}; 0),
\end{equation}
with the slab-truncated Coulomb interaction included \cite{sor}, 
\begin{equation}
	v_{\rm t}^{\rm slab}({\rm \bf{q}}) = \frac{4\pi}{q^2}\cdot
	\left( 1-e^{-q_{xy}\cdot z_{c}} {\rm cos}(q_z\cdot z_c)\right),
\end{equation}
where $z_c$ is the truncation distance in the perpendicular direction. Such kind of treatment could guarantee the convergence of dielectric screening (``head'' in $\varepsilon_{\rm \bf{GG'}}$) to approach to unit in the long wavelength limit.

Within the generalized plasmon pole (GPP) model, the imaginary and real part inverse 
dielectric matrix with finite frequencies are given by \cite{gw}
\begin{equation}
	{\rm Im}\varepsilon_{\rm \bf{GG'}}^{-1}({\rm\bf{q}},\omega) = 
	-\frac{\pi}{2}\frac{\Omega_{\rm \bf{GG'}}({\rm \bf{q}})}{\tilde{\omega}_{\rm \bf{GG'}}({\rm \bf{q}})}
	\left\{\delta[\omega-\tilde{\omega}_{\rm \bf{GG'}}({\rm \bf{q}})]-\delta[\omega+\tilde{\omega}_{\rm \bf{GG'}}({\rm \bf{q}})]\right\},
\end{equation}
and 
\begin{equation}
	{\rm Re}\varepsilon_{\rm \bf{GG'}}^{-1}({\rm\bf{q}},\omega) = 1+\frac{\Omega_{\rm \bf{GG'}}^2({\rm \bf{q}})}
	{\omega^2-\tilde{\omega}_{\rm \bf{GG'}}^2({\rm \bf{q}})},
\end{equation}
where $\Omega_{\rm \bf{GG'}}({\rm \bf{q}})$ and $\tilde{\omega}_{\rm \bf{GG'}}({\rm \bf{q}})$ are the effective 
bare plasma frequency and the GPP mode frequency defined as \cite{gw}: 
\begin{equation}
	\Omega^2_{\rm \bf{GG'}}({\rm \bf{q}}) = \omega_{\rm p}^2
	\frac{({\rm \bf{q}}+{\rm \bf{G}})\cdot({\rm \bf{q}}+{\rm \bf{G'}})}{|{\rm \bf{q}}+{\rm \bf{G}}|^2}
	\frac{\rho({\rm \bf{G}}-{\rm \bf{G'}})}{\rho(\bf{0})},
\end{equation}
\begin{equation}
	\tilde{\omega}_{\rm \bf{GG'}}^2({\rm \bf{q}})=
	\frac{\Omega_{\rm \bf{GG'}}^2({\rm \bf{q}})}{\delta_{\rm \bf{GG'}}-\varepsilon_{\rm \bf{GG'}}^{-1}({\rm\bf{q}},\omega=0)}.
\end{equation}
Here $\rho$ is the electron charge density in reciprocal space and $\omega_{\rm p}^2 = 4\pi\rho({\bf 0})e^2/m$ is the 
classical plasma frequency. 

Using the form of the above frequency-dependent dielectric function, the self-energy operator $\rm \Sigma$ is solved 
in two parts, $\rm \Sigma = \Sigma_{SEX} + \Sigma_{COH}$, where $\rm \Sigma_{SEX}$ is the screened exchange operator 
and $\rm \Sigma_{COH}$ is the Coulomb-hole operator as \cite{gw}
\begin{equation}
	\begin{aligned}
		\langle n,{\rm \bf{k}} | {\rm \Sigma_{SEX}}(E) | n',{\rm \bf{k}} \rangle = 
		-\sum_{n_1}^{\rm occ}\sum_{\rm \bf{q,G,G'}}
		\langle n,{\rm \bf{k}} | e^{i({\rm \bf{q}}+{\rm \bf{G}})\cdot{\rm \bf{r}}} | n_1,{\rm \bf{k}} - {\rm \bf{q}} \rangle
		\langle n_1,{\rm \bf{k}} - {\rm \bf{q}} | e^{-i({\rm \bf{q}}+{\rm \bf{G'}})\cdot{\rm \bf{r'}}} | n',{\rm \bf{k}} \rangle
		\\
		\times
		\left( 
		1+\frac{\Omega_{\rm \bf{GG'}}^2({\rm \bf{q}})}
		{(E-E_{n_1{\rm \bf{k-q}}})^2-\tilde{\omega}_{\rm \bf{GG'}}^2({\rm \bf{q}})}
		\right)
		v({\rm \bf{q+G'}}),
	\end{aligned}
\end{equation}
and
\begin{equation}
	\begin{aligned}
		\langle n,{\rm \bf{k}} | {\rm \Sigma_{COH}}(E) | n',{\rm \bf{k}} \rangle = 
		\sum_{n_1}\sum_{\rm \bf{q,G,G'}}
		\langle n,{\rm \bf{k}} | e^{i({\rm \bf{q}}+{\rm \bf{G}})\cdot{\rm \bf{r}}} | n_1,{\rm \bf{k}} - {\rm \bf{q}} \rangle
		\langle n_1,{\rm \bf{k}} - {\rm \bf{q}} | e^{-i({\rm \bf{q}}+{\rm \bf{G'}})\cdot{\rm \bf{r'}}} | n',{\rm \bf{k}} \rangle
		\\
		\times
		\frac{1}{2}\frac{\Omega_{\rm \bf{GG'}}^2({\rm \bf{q}})}
		{\tilde{\omega}_{\rm \bf{GG'}}({\rm \bf{q}})[E-E_{n_1{\rm \bf{k-q}}}-\tilde{\omega}_{\rm \bf{GG'}}({\rm \bf{q}})]}v({\rm \bf{q+G'}}).
	\end{aligned}
\end{equation} 

With the above obtained quasiparticle energies and static dielectric screening from RPA, the electron-hole excitations are then calculated by solving the
BSE for each exciton state \emph{S} \cite{bse}: 
\begin{equation}
	\left( E^{\rm QP}_{c{\rm \bf{k}}}-E^{\rm QP}_{v{\rm \bf{k}}}
	\right)A^{S}_{vc{\rm \bf{k}}} + \sum_{v'c'{\rm \bf{k}}'}\left
	\langle vc{\rm \bf{k}}\left| K^{\rm eh} \right|v'c'{\rm \bf{k}}'
	\right \rangle A^{S}_{v'c'{\rm \bf{k'}}} = {\Omega}^{S}A^{S}_{vc{\rm \bf{k}}},
\end{equation} where $A^S_{vc{\rm \bf{k}}}$ is the exciton
wavefunction, $\Omega^S$ is the excitation energy, and $K^{\rm
	eh}$ is the electron-hole interaction kernel. 

The kernel contains two terms, a screened direct interaction and a bare exchange interaction, 
$K^{\rm eh} = K^{\rm d} + K^{\rm x}$, defined as \cite{bse}:
\begin{equation}
	\left \langle vc{\rm \bf{k}}\left| K^{\rm d} \right|v'c'{\rm \bf{k}}' \right \rangle = 
	\sum_{\rm \bf{GG'}}
	\langle c,{\rm \bf{k+q}} | e^{-i({\rm \bf{q}}+{\rm \bf{G}})\cdot{\rm \bf{r}}} | c',{\rm \bf{k}} \rangle
	W_{\rm \bf{GG'}}({\rm \bf{q}}; 0)
	\langle v',{\rm \bf{k}} | e^{i({\rm \bf{q}}+{\rm \bf{G'}})\cdot{\rm \bf{r}}} | v,{\rm \bf{k+q}} \rangle,
\end{equation}
and
\begin{equation}
	\left \langle vc{\rm \bf{k}}\left| K^{\rm x} \right|v'c'{\rm \bf{k}}' \right \rangle = 
	\sum_{\rm \bf{GG'}}
	\langle c,{\rm \bf{k+q}} | e^{-i({\rm \bf{q}}+{\rm \bf{G}})\cdot{\rm \bf{r}}} | v,{\rm \bf{k}} \rangle
	\delta_{\rm \bf{GG'}}v({\rm \bf{q}}+{\rm \bf{G}})
	\langle v',{\rm \bf{k}} | e^{i({\rm \bf{q}}+{\rm \bf{G'}})\cdot{\rm \bf{r}}} | c',{\rm \bf{k+q}} \rangle.
\end{equation}

Finally, we obtain
the imaginary parts of frequency-dependent complex dielectric
function $\varepsilon_2(\omega)$ as \cite{bse}
\begin{equation}
	\varepsilon_2(\omega)=\frac{16\pi^2e^2}{\omega^2}\sum_S\left|{\rm
		\bf{e}}\cdot\left \langle0|{\rm \bf{v}}|S\right
	\rangle\right|^2\delta\left( \omega-\Omega^S \right),
\end{equation} where $\rm \bf{v}$ is the velocity operator and $\rm
\bf{e}$ is the polarization of the incoming light. Here, we use
absorbance $\mathcal{A}$ of 2D materials to measure their optical
properties, which is expressed as \cite{manos-gese}
\begin{equation}
	\mathcal{A}(\omega)=1-e^{-\alpha(\omega)
		d}=1-e^{-\frac{\omega\varepsilon_{2}d}{\hbar c}}, \end{equation}
where $\alpha$ is the absorption coefficient, $\textit{d}$ is the
thickness of the simulation cell along the direction perpendicular
to the layer, $\varepsilon_2$ is the imaginary part of the dielectric
function, and $\omega$ is the photon energy, respectively. 

\section{CONVERGENCE OF $G_0W_0$-BSE CALCULATIONS OF CrSBr MONOLAYER}

In Fig. 12, we show the convergence of quasiparticle energies
in CrSBr with respect to (a) the scale of $k$ grid (coarse
grid), (b) the number of bands, and (c) the dielectric cutoff. It
is noted that fine sampling is necessary to capture the rapid
variation in screening at small wave vectors and the fine
features in the exciton wave functions, which are tightly
localized in \textit{k} space. In Fig. 13, besides the RPA dielectric function, we show
the convergence of exciton energy in CrSBr with respect
to the scale of $k$ grid (fine grid).

\begin{figure}[H]
\centering
	\includegraphics[width=0.45\textwidth]{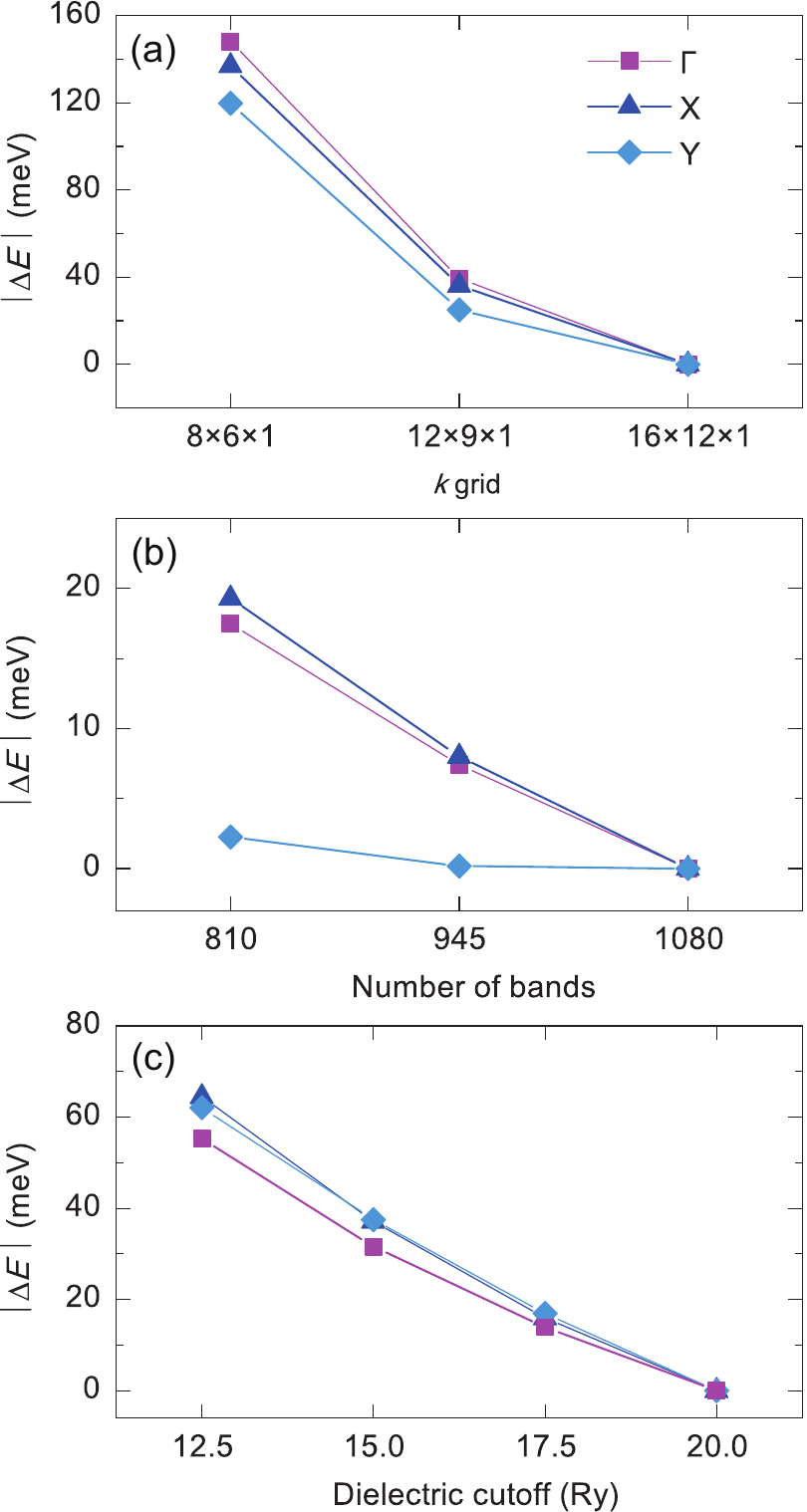}
	\caption{\label{fig:12} Convergence of the quasiparticle
		(\emph{$G_0W_0$}) band gap at three high symmetry $k$ points with
		respect to the scale of coarse \emph{k} grid (a), number of
		bands (b), and dielectric cutoff energy ($\varepsilon_{\rm cutoff}$) (c). In (a), we use a $\varepsilon_{\rm cutoff}$ of 20 Ry
		and 1080 bands; in (b), we use a \emph{k} grid of $16 \times 12
		\times 1$ and 1080 bands; and in (c), we use a $\varepsilon_{\rm
			cutoff}$ of 20 Ry and a \emph{k} grid of $16\times12\times1$. With
		the above consideration, we use a $16\times12\times1$ \emph{k}
		grid, a $\varepsilon_{\rm cutoff}$ of 20 Ry, and 1080 (20 times larger) bands for our
		$G_0W_0$ calculations.}
\end{figure}

\begin{figure}[H]
\centering
	\includegraphics[width=0.45\textwidth]{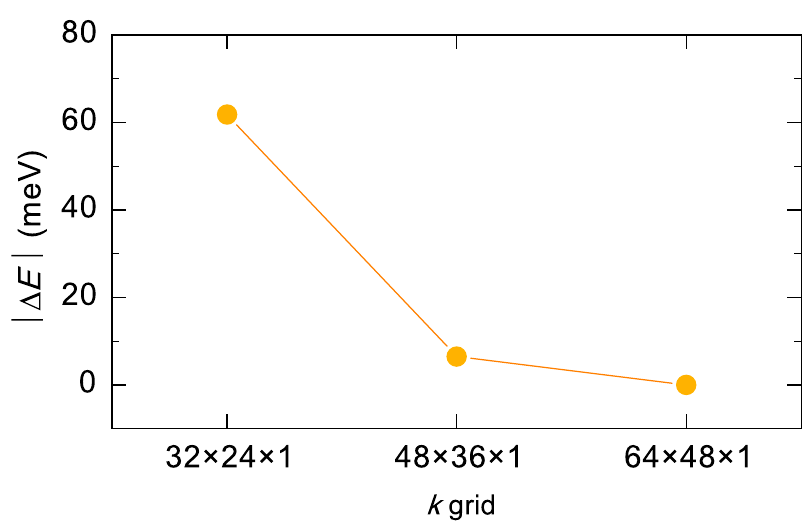}
	\caption{\label{fig:13} Convergence of the energy of the
		1\emph{s} exciton state in BSE calculations with respect to the
		scale of the fine \emph{k} grids: including $32\times24\times1$,
		$60\times60\times1$, $48\times36\times1$, and
		$64\times48\times1$. With the above consideration, we use a
		$64\times48\times1$ \emph{k} grid in our BSE calculations.}
\end{figure}

\section{MAXIMUM LOCALIZED WANNIER ORBITALS IN CRSBR}

Using Wannier90 \cite{w90}, the orbital-decomposed band
structure (PBE level) is shown in Fig. 14.

\begin{figure}
	\includegraphics[width=0.45\textwidth]{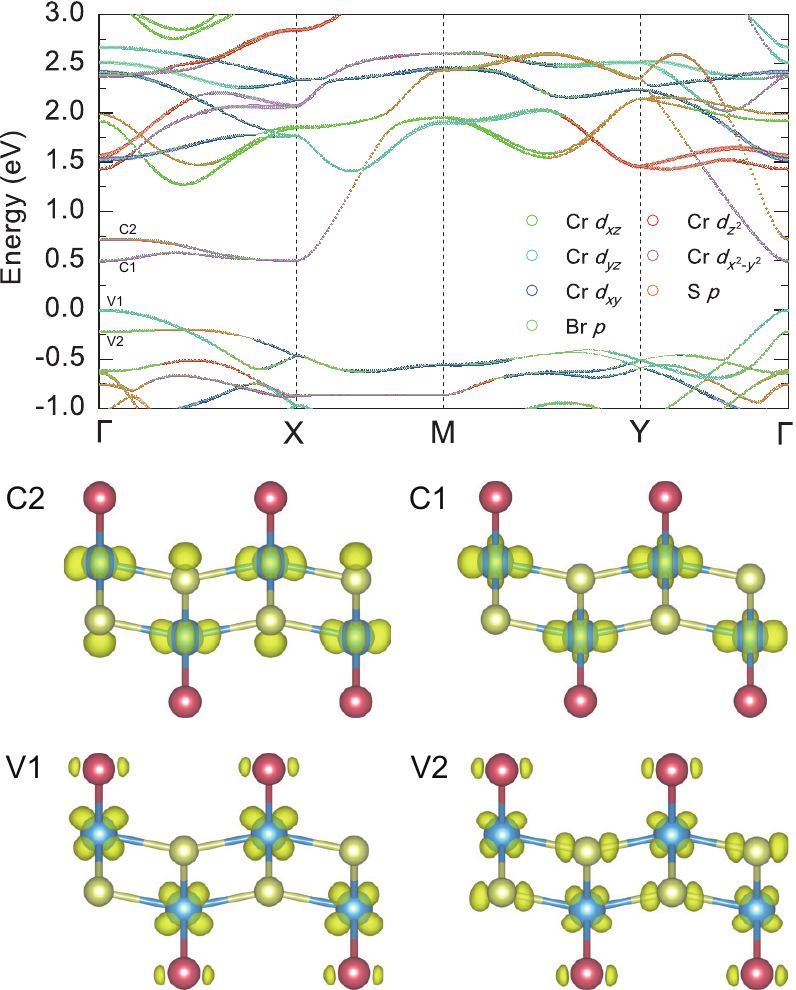}
	\caption{\label{fig:14} The up panel shows the orbital-decomposed
		band structure (DFT-PBE level) of CrSBr from maximum
		localized Wannier orbitals using Wannier90 \cite{w90}. The partial charge at $\Gamma$ point
		is shown in the down panel.}
\end{figure}

\section{CALCULATIONS OF LINEAR DICHROISM (OPTICAL BIREFRIGENCE) AND SCHAFER-HUBERT AND VOIGT EFFECTS}

As shown in Fig. 1, considering the in-plane magnetization along \textit{y} axis ($\bm{B}=B\bm{\hat{e}_{y}}$), the dielectric tensor is expressed as:
\begin{equation}
	\bm{\varepsilon}(\omega,\bm{B}) =
	\begin{pmatrix}
		\varepsilon_{xx}(\omega,\bm{B}) & 0 & \varepsilon_{xz}(\omega,\bm{B}) \\
		0 & \varepsilon_{yy}(\omega,\bm{B}) & 0 \\
		-\varepsilon_{xz}(\omega,\bm{B}) & 0 & \varepsilon_{zz}(\omega,\bm{B}) \\
	\end{pmatrix}.
\end{equation} The Fresnel equation for the propagation of electromagnetic wave is given by:
\begin{equation}
	[n^{2}\bm{I}-\bm{\varepsilon}-\bm{n}:\bm{n}]\cdot\bm{E}=0.
\end{equation}
For normal incidence, the complex refractory index \textbf{n} is:
\begin{equation}
	\bm{n}=  \frac{c\bm{k}}{\omega}=\frac{ck}{\omega}\bm{\hat{e}_{z}}.
\end{equation}
By solving the above Fresnel equations, we 
get the normal modes as the $\parallel$ and $\perp$ for linearly polarized plane waves which are parallel and perpendicular to $\bm{\hat{e}_{y}}$ (magnetization direction), with distinct refractive indices:
\begin{equation}
	n^{2}_{\parallel}(\omega,B\bm{\hat{e}_{y}})=\varepsilon_{yy}(\omega,B\bm{\hat{e}_{y}}),
\end{equation}
and
\begin{equation}
	n^{2}_{\perp}(\omega,B\bm{\hat{e}_{y}})=\varepsilon_{xx}(\omega,B\bm{\hat{e}_{y}})+\frac{\varepsilon_{xz}^{2}(\omega,B\bm{\hat{e}_{y}})}{\varepsilon_{zz}(\omega,B\bm{\hat{e}_{y}})}.
\end{equation}

To mathematically describe how an electromagnetic wave interacts with such stratified and anisotropic media, we adopt a 4x4 formalism involving the in-plane components of both
electric $(E_{x}, E_{y})$ and magnetic fields $(B_{x}, B_{y})$. We consider a monolayer material
magnetized along the $+y$ direction, and the second layer is
vacuum. Within each layer ($l$ = 0, 1, 2), we choose the four
eigenmodes of light as follows:$\,\bm{\hat{e}^{(l)}_{1}}=\bm{\hat{e}^{(l)}_{2}}=\bm{\hat{e}_{x}}, \,\bm{\hat{e}^{(l)}_{3}}=\bm{\hat{e}^{(l)}_{4}}=\bm{\hat{e}_{y}}$, with the corresponding refractive indices:$\,n^{(l)}_{1}=-n^{(l)}_{2}=n^{(l)}_{\perp},\,n^{(l)}_{3}=-n^{(l)}_{4}=n^{(l)}_{\parallel}$. The electric and magnetic fields of
light in the first and second layers are given by:
\begin{equation}
	\bm{E}^{(l)}=\sum_{j=1}^{4}E^{(l)}_{0j}\bm{\hat{e}}^{(l)}_{j}\exp \{i[k^{(l)}_{j}(z-z_{l-1})-\omega t] \},
\end{equation}
and
\begin{equation}
	c\bm{B}^{(l)}=\sum_{j=1}^{4}E^{(l)}_{0j}\bm{b}^{(l)}_{j}\exp \{i[k^{(l)}_{j}(z-z_{l-1})-\omega t] \},
\end{equation}
with $\bm{b}^{(l)}_{j} = n^{(l)}_{j}\bm{\hat{e}}_{z} \times \bm{\hat{e}}^{(l)}_{j}$. The electric and magnetic fields of light in the zeroth layer are given by,
\begin{equation}
	\bm{E}^{(0)}=\sum_{j=1}^{4}E^{(0)}_{0j}\bm{\hat{e}}^{(0)}_{j}\exp \{i[k^{(0)}_{j}(z-z_{0})-\omega t] \},
\end{equation}
and
\begin{equation}
	c\bm{B}^{(0)}=\sum_{j=1}^{4}E^{(0)}_{0j}\bm{b}^{(0)}_{j}\exp \{i[k^{(0)}_{j}(z-z_{0})-\omega t] \}.
\end{equation}
The requirement of the continuity of the tangential field
components at the interfaces connects the field amplitudes $E^{(l)}_{0j}$ between two neighboring layers. The dynamical matrix within each layer
is given by a block-diagonal form:
\begin{equation}
	\begin{aligned}
		\bm{D}^{(l)} &=
		\begin{pmatrix}
			\bm{\hat{e}}^{(l)}_{1} \cdot \bm{\hat{e}}_{x} & \bm{\hat{e}}^{(l)}_{2}\cdot \bm{\hat{e}}_{x} & \bm{\hat{e}}^{(l)}_{3}\cdot \bm{\hat{e}}_{x} &\bm{\hat{e}}^{(l)}_{4}\cdot \bm{\hat{e}}_{x}\\
			\bm{\hat{b}}^{(l)}_{1}\cdot \bm{\hat{e}}_{y} & \bm{\hat{b}}^{(l)}_{2}\cdot \bm{\hat{e}}_{y} & \bm{\hat{b}}^{(l)}_{3}\cdot \bm{\hat{e}}_{y} &\bm{\hat{b}}^{(l)}_{4}\cdot \bm{\hat{e}}_{y}\\
			\bm{\hat{e}}^{(l)}_{1}\cdot \bm{\hat{e}}_{y} & \bm{\hat{e}}^{(l)}_{2}\cdot \bm{\hat{e}}_{y} & \bm{\hat{e}}^{(l)}_{3}\cdot \bm{\hat{e}}_{y} &\bm{\hat{e}}^{(l)}_{4}\cdot \bm{\hat{e}}_{y}\\
			\bm{\hat{b}}^{(l)}_{1}\cdot \bm{\hat{e}}_{x} & \bm{\hat{b}}^{(l)}_{2}\cdot \bm{\hat{e}}_{x} & \bm{\hat{b}}^{(l)}_{3}\cdot \bm{\hat{e}}_{x} &\bm{\hat{b}}^{(l)}_{4}\cdot \bm{\hat{e}}_{x}\\
		\end{pmatrix} \\
		&=
		\begin{pmatrix}
			1 & 1 & 0 &0\\
			n^{(l)}_{\perp} & -n^{(l)}_{\perp} & 0 &0\\
			0 & 0 & 1 &1\\
			0 & 0 & -n^{(l)}_{\parallel} &n^{(l)}_{\parallel}\\
		\end{pmatrix}.
	\end{aligned}
\end{equation}
The propagation matrix is defined as a diagonal matrix:
\begin{equation}
	\bm{P}^{(1)}=
	\begin{pmatrix}
		e^{i\delta_{\perp}} & 0 & 0 &0\\
		0 & e^{-i\delta_{\perp}} & 0 &0\\
		0 & 0 & e^{i\delta_{\parallel}} &0\\
		0 & 0 & 0 &e^{-i\delta_{\parallel}}\\
	\end{pmatrix},
\end{equation}
where $\delta_{\perp}=\frac{\omega}{c}n_{\perp}$, $\delta_{\parallel}=\frac{\omega}{c}n_{\parallel}$, and $d$ is the thickness of layer 1, i.e. monolayer CrSBr. 

In this two-interface setup, $\bm{E}^{(0)}_{0}$
and $\bm{E}^{(2)}_{0}$ are related by the transfer matrix $\bm{M}_{c}$ in the basis of
linearly polarized lights:
\begin{equation}
	\bm{E}^{(2)}_{0}=\bm{M}_{c}\bm{E}^{(0)}_{0}=[\bm{D}^{(2)}]^{-1}\bm{D}^{(1)}\bm{P}^{(1)}[\bm{D}^{(1)}]^{-1}\bm{D}^{(0)}\bm{E}^{(0)}_{0}.
\end{equation}
$\bm{M}_{c}$ has a simple block-diagonal form:
\begin{equation}
	\bm{M}_{c}=
	\begin{pmatrix}
		\bm{M}_{\perp} & 0 \\
		0 & \bm{M}_{\parallel} \\
	\end{pmatrix},
\end{equation}
\begin{equation}
	\bm{M}_{\perp}= \frac{1}{t^{\perp}_{21}t^{\perp}_{10}}
	\begin{pmatrix}
		e^{i\delta_{\perp}}+e^{-i\delta_{\perp}}r^{\perp}_{21}r^{\perp}_{10} & e^{i\delta_{\perp}}r^{\perp}_{10}+e^{-i\delta_{\perp}}r^{\perp}_{21} \\
		e^{i\delta_{\perp}}r^{\perp}_{21}+e^{-i\delta_{\perp}}r^{\perp}_{10} & e^{i\delta_{\perp}}r^{\perp}_{21}r^{\perp}_{10}+e^{-i\delta_{\perp}} \\
	\end{pmatrix},
\end{equation}
where $r_{mn}$ and $t_{mn}$ are the Fresnel coefficients of the interface
from the \textit{m}th layer to the \textit{n}th layer. And $\bm{M}_{\parallel}$ has the same form.
In the left and right layers, we adopt a basis transformation from the linearly polarized light ($\bm{\hat{e}}_{x}$,$\bm{\hat{e}}_{y}$) to the light ($\bm{\hat{e}}_{a}$,$\bm{\hat{e}}_{b}$) with $\bm{\hat{e}}_{a}=\frac{1}{\sqrt{2}}(\bm{\hat{e}}_{x}+\bm{\hat{e}}_{y})$ and $\bm{\hat{e}}_{b}=\frac{1}{\sqrt{2}}(-\bm{\hat{e}}_{x}+\bm{\hat{e}}_{y})$. $\bm{\hat{e}}^{(l)}_{1}=\bm{\hat{e}}^{(l)}_{2}=\bm{\hat{e}}_{a}$, $\bm{\hat{e}}^{(l)}_{3}=\bm{\hat{e}}^{(l)}_{4}=\bm{\hat{e}}_{b}$, and $n^{(l)}_{1}=-n^{(l)}_{2}=n^{(l)}_{3}=-n^{(l)}_{4}=n^{(l)}$, for $l$ $=$ $0$, $2$. This new basis of linearly
polarized plane waves is denoted as $\{a\rightarrow,a\leftarrow,b\rightarrow,b\leftarrow\}$. In this basis of linearly polarized lights, the electric field amplitudes in the left and right layer are related by transfer matrix $M$:
\begin{equation}
	\begin{pmatrix}
		E^{(2)}_{0a\rightarrow} \\
		E^{(2)}_{0a\leftarrow} \\
		E^{(2)}_{0b\rightarrow} \\
		E^{(2)}_{0b\leftarrow} \\
	\end{pmatrix}
	=\bm{M}
	\begin{pmatrix}
		E^{(0)}_{0a\rightarrow} \\
		E^{(0)}_{0a\leftarrow} \\
		E^{(0)}_{0b\rightarrow} \\
		E^{(0)}_{0b\leftarrow} \\
	\end{pmatrix}
\end{equation}
\begin{equation}
	\bm{M}=\frac{1}{2}
	\begin{pmatrix}
		\bm{M}_{\perp}+\bm{M}_{\parallel} & -\bm{M}_{\perp}+\bm{M}_{\parallel} \\
		-\bm{M}_{\perp}+\bm{M}_{\parallel} & \bm{M}_{\perp}+\bm{M}_{\parallel} \\
	\end{pmatrix}.
\end{equation}

Because the angle of electric field of incident light and the direction
of magnetization ($+y$) is $45^{\circ}$, $E^{(2)}_{0b\leftarrow}=0$. In addition, there are no reflecting lights from the
\textit{zero}th medium (semi-infinite substrate) to CrSBr, which means $E^{(0)}_{0a\rightarrow}=E^{(0)}_{0b\rightarrow}=0$. With these two conditions, by calculating Eq. D15, we can get:
\begin{equation}
	t_{ss}\equiv\frac{E^{(0)}_{0a\leftarrow}}{E^{(2)}_{0a\leftarrow}}=\frac{\bm{M}_{44}}{\bm{M}_{22}\bm{M}_{44}-\bm{M}_{24}\bm{M}_{42}}
\end{equation}
\begin{equation}		t_{sp}\equiv\frac{E^{(0)}_{0b\leftarrow}}{E^{(2)}_{0a\leftarrow}}=\frac{-\bm{M}_{42}}{\bm{M}_{22}\bm{M}_{44}-\bm{M}_{24}\bm{M}_{42}}
\end{equation}
\begin{equation}
	r_{ss}\equiv\frac{E^{(0)}_{0a\rightarrow}}{E^{(2)}_{0a\leftarrow}}=\frac{\bm{M}_{12}\bm{M}_{44}-\bm{M}_{14}\bm{M}_{42}}{\bm{M}_{22}\bm{M}_{44}-\bm{M}_{24}\bm{M}_{42}}
\end{equation}
\begin{equation}
	r_{sp}\equiv\frac{E^{(0)}_{0b\rightarrow}}{E^{(2)}_{0a\leftarrow}}=\frac{\bm{M}_{32}\bm{M}_{44}-\bm{M}_{34}\bm{M}_{42}}{\bm{M}_{22}\bm{M}_{44}-\bm{M}_{24}\bm{M}_{42}}
\end{equation}
The SH signals for the reflected (right-moving)
electric field $E^{(2)}_{\rightarrow}=E^{(2)}_{0a\rightarrow}\bm{\hat{e}}_{a}+E^{(2)}_{0b\rightarrow}\bm{\hat{e}}_{b}$ are expressed as:
\begin{equation}
	\tan2\theta_{\rm{SH}}= \frac{2|\frac{r_{sp}}{r_{ss}}|\cos(\arg\frac{r_{sp}}{r_{ss}})}{1-|\frac{r_{sp}}{r_{ss}}|^{2}},
\end{equation}
\begin{equation}
	\sin2\eta_{\rm{SH}}= \frac{2|\frac{r_{sp}}{r_{ss}}|\sin(\arg\frac{r_{sp}}{r_{ss}})}{1+|\frac{r_{sp}}{r_{ss}}|^{2}},
\end{equation}
Similarly, the Voigt signals for the transmitted (left-moving)
electric field
$E^{(0)}_{\leftarrow}=E^{(0)}_{0a\leftarrow}\bm{\hat{e}}_{a}+E^{(0)}_{0b\leftarrow}\bm{\hat{e}}_{b}$ are expressed as:
\begin{equation}
	\tan2\theta_{\rm{V}}= \frac{2|\frac{t_{sp}}{t_{ss}}|\cos(\arg\frac{t_{sp}}{t_{ss}})}{1-|\frac{t_{sp}}{t_{ss}}|^{2}},
\end{equation}
\begin{equation}
	\sin2\eta_{\rm{V}}= \frac{2|\frac{t_{sp}}{t_{ss}}|\sin(\arg\frac{t_{sp}}{t_{ss}})}{1+|\frac{t_{sp}}{t_{ss}}|^{2}}.
\end{equation}

In this work, we rescale the calculated dielectric function in a slab model by the thickness of a monolayer material:
\begin{equation}
	\varepsilon_{\alpha\alpha}=1+\frac{l}{d}(\tilde{\varepsilon}_{\alpha\alpha}-1),
\end{equation}
and
\begin{equation}
	\varepsilon_{\alpha\beta}=\frac{l}{d}\tilde{\varepsilon}_{\alpha\beta},
\end{equation}
where $l$ and $d$ are  thicknesses of the slab model along the out-of-plane direction and monolayer material, and $\tilde{\varepsilon}_{\alpha\alpha}$ and $\tilde{\varepsilon}_{\alpha\beta}$ are calculated dielectric functions in the BerkeleyGW package. 

$\tilde{\varepsilon}_{\alpha\alpha}$ and $\tilde{\varepsilon}_{\alpha\beta}$ can be expressed as follows:
\begin{equation}
	{\rm Im}[\varepsilon_{\alpha\alpha}(\omega)]= \frac{\pi\hbar^{2}}{\varepsilon_{0}N_{k}V}\sum_{S}\frac{1}{\Omega^{2}_{S}}|\langle0|\hat{j}^{\alpha}_{p}|S\rangle|^{2}\left[\delta(\hbar\omega-\Omega_{S})\right],
\end{equation}
\begin{equation}
	{\rm Re}[\varepsilon_{\alpha\alpha}(\omega)]= -\frac{\hbar^{2}}{\varepsilon_{0}N_{k}V}\sum_{S}\frac{1}{\Omega^{2}_{S}}|\langle0|\hat{j}^{\alpha}_{p}|S\rangle|^{2}\left[\frac{1}{\hbar\omega-\Omega_{S}}\right],
\end{equation}
and
\begin{equation}
	{\rm Im}[\varepsilon_{\alpha\beta}(\omega)]= \frac{i\hbar^{2}}{\varepsilon_{0}N_{k}V}\sum_{S}\frac{1}{\Omega^{2}_{S}}\left[\frac{\langle0|\hat{j}^{\alpha}_{p}|S\rangle\langle S|\hat{j}^{\beta}_{p}|0\rangle}{\hbar\omega-\Omega_{S}}\right],
\end{equation}
\begin{equation}
	\begin{aligned}
		{\rm Re}[\varepsilon_{\alpha\beta}(\omega)] = \frac{i\pi\hbar^{2}}{\varepsilon_{0}N_{k}V}&\sum_{S}\frac{1}{\Omega^{2}_{S}}[\langle0|\hat{j}^{\alpha}_{p}|S\rangle\langle S|\hat{j}^{\beta}_{p}|0\rangle\delta(\hbar\omega-\Omega_{S})],
	\end{aligned}
\end{equation}
where $N_{k}$ is the number of k-points, $V$ is the volume of a unit cell,  $\bm{\hat{j}_{p}}=-e\bm{\hat{v}}$, and $\langle0|\hat{j}^{\alpha}_{p}|S\rangle=\sum_{cv\bm{k}}A^{S}_{cv\bm{k}}\langle v\bm{k}|\hat{j}^{\alpha}_{p}|c\bm{k}\rangle$.

\section{CALCULATIONS OF SH AND VOIGT EFFECTS IN CRSBR}

In Fig. 15, we have shown the calculated results without the consideration of electron-hole interaction, i.e., under independent particle approximation. In Fig. 16, we have compared the results with and without the inclusion of off-diagonal components of dielectric function from BSE.

\begin{figure}[H]
	\centering
	\includegraphics[width=0.5\textwidth]{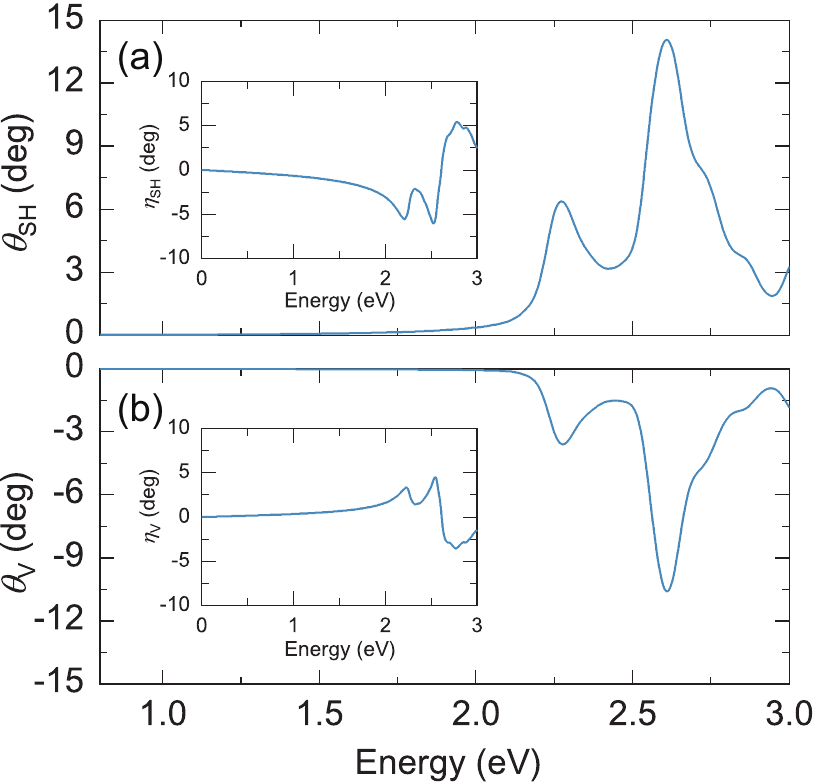}
	\caption{\label{fig:15} Rotation angles of (a) reflected and (b) transmitted light in CrSBr monolayer without the inclusion of electron-hole interactions. Insets are the corresponding ellipticities.}
\end{figure}

\begin{figure}[H]
	\centering
	\includegraphics[width=0.45\textwidth]{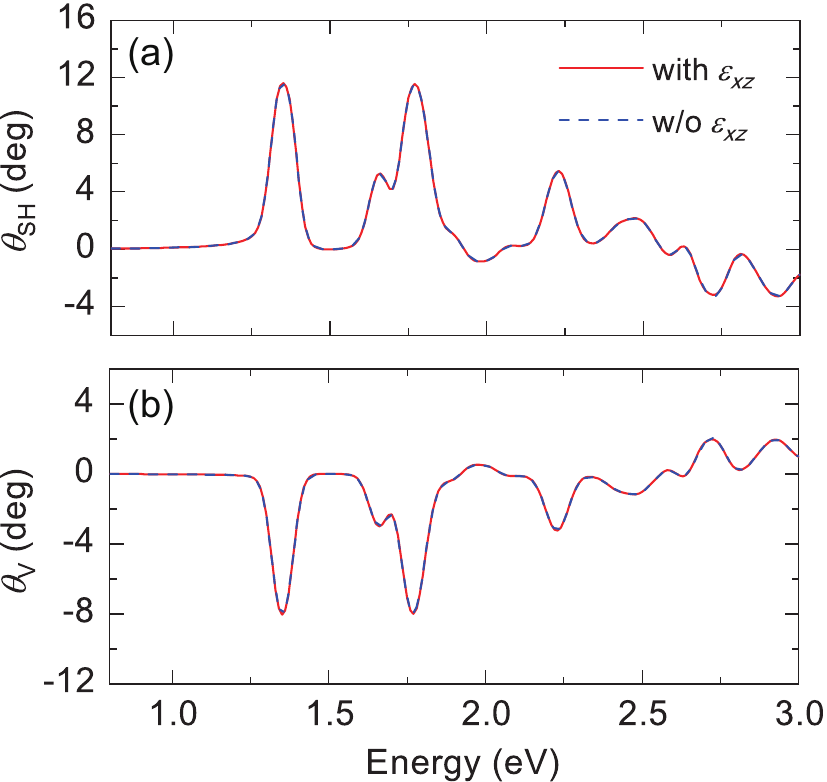}
	\caption{\label{fig:16} Comparison of rotation angles for both (a) reflected and (b) transmitted lights in CrSBr monolayer between the situations with and without the inclusion of off-diagonal component $\varepsilon_{xz}$.}
\end{figure}

\section{ABSORBANCE AND EXCITON SPECTRUM WITH SPIN ORIENTATION ALONG X AND Z AXIS}
In Fig. 17 and Fig. 18, we have shown the anisotropic optical absorbance and exciton spectrum in CrSBr monolayer for states with spin orientation along $x$ axis and $z$ axis, respectively. Optical absorbance and exciton spectrum for the state with spin orientation along $x$ axis are similar with those for the state with spin orientation along $z$ axis.
\begin{figure}[H]
	\centering
	\includegraphics[width=0.45\textwidth]{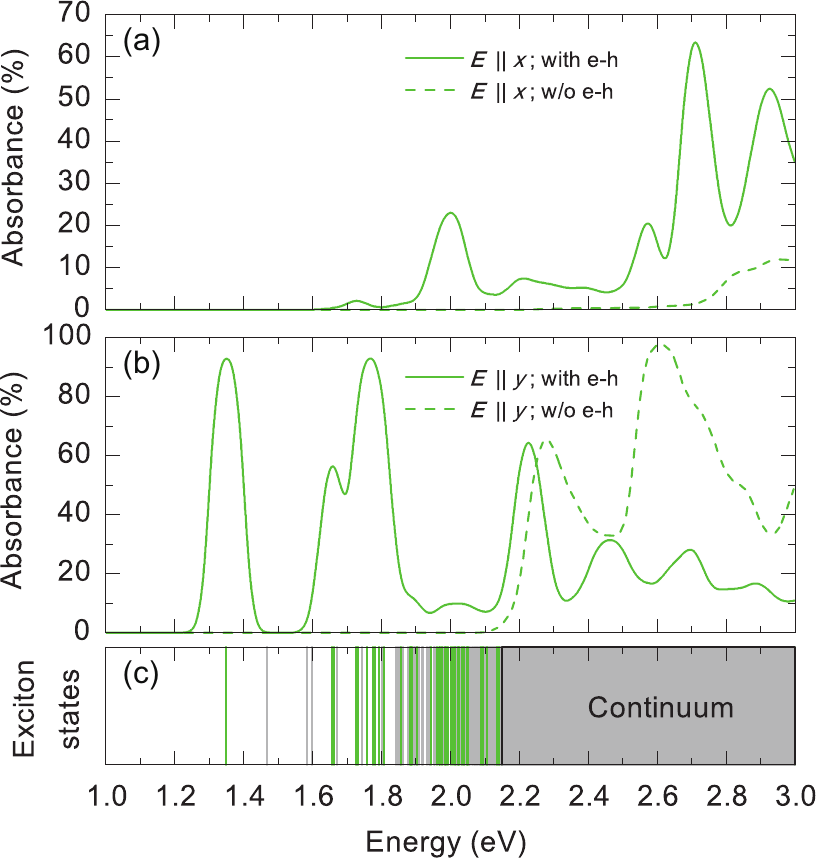}
	\caption{\label{fig:17} Anisotropic optical absorbance in CrSBr monolayer with spin orientation along $x$ axis for the electric polarization along \textit{x} (a) and \textit{y} (b) directions, respectively. (c) Exciton spectrum.}
\end{figure}

\begin{figure}[H]
	\centering
	\includegraphics[width=0.45\textwidth]{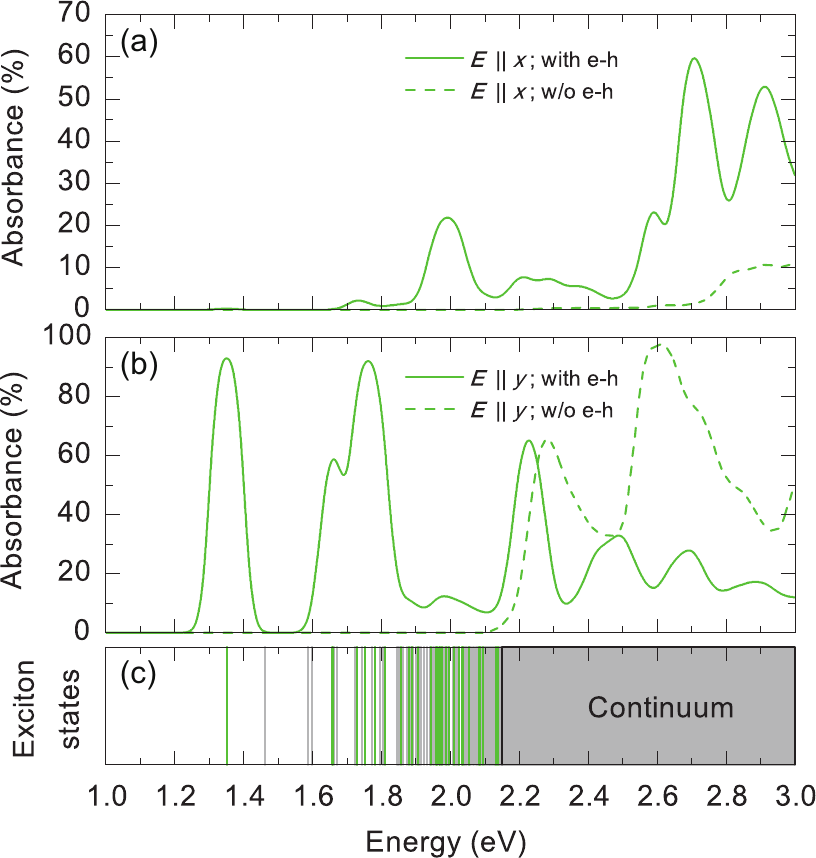}
	\caption{\label{fig:18} Anisotropic optical absorbance in CrSBr monolayer with spin orientation along $z$ for the electric polarization along \textit{x} (a) and \textit{y} (b) directions, respectively. (c) Exciton spectrum.}
\end{figure}

\section{COMPARISON OF ROTATION ANGLES OF SH AND VOIGT EFFECTS FOR THE STATES WITH SPIN ORIENTATION ALONG X AND Y AXIS}
In Fig. 19, we have compared rotation angles for states with spin orientation along $x$ and $y$ axis. It is obvious that rotation angles for the state with spin orientation along $x$ axis are nearly unchanged compared with those for the state with spin orientation along $y$ axis.

\begin{figure}[H]
	\centering
	\includegraphics[width=0.45\textwidth]{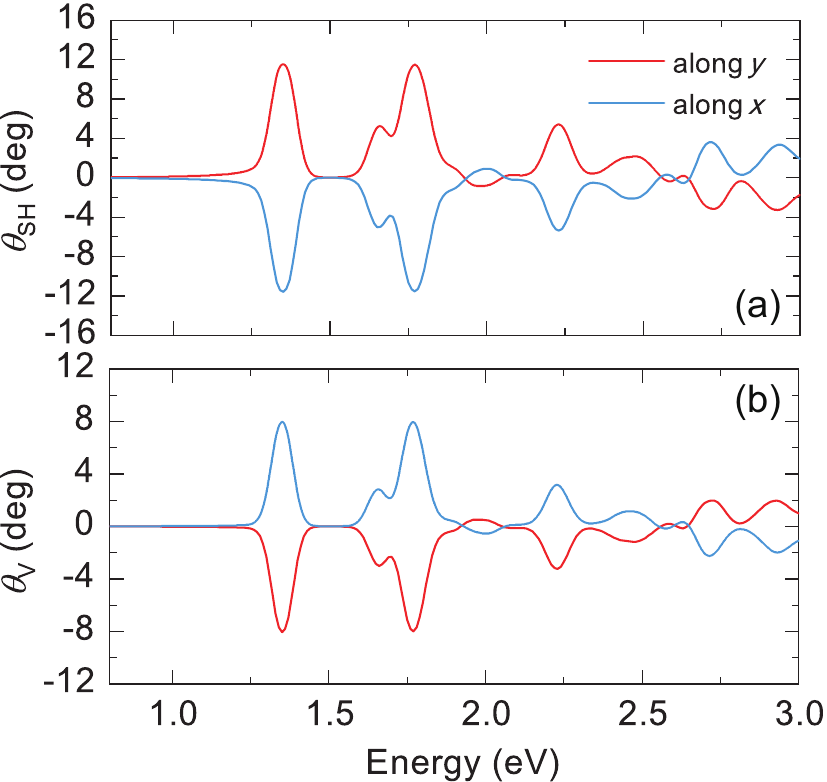}
	\caption{\label{fig:19} Rotation angles of (a) reflected and (b) transmitted light in CrSBr monolayer for states with spin orientation along $x$ (blue) and $y$ (red) axis considering electron-hole interactions.}
\end{figure}

\section{CALCULATIONS OF KERR AND FARADAY EFFECTS IN CRSBR WITH OUT-OF-PLANE MAGNETIZATION}

\begin{figure}[H]
	\centering
	\includegraphics[width=0.5\textwidth]{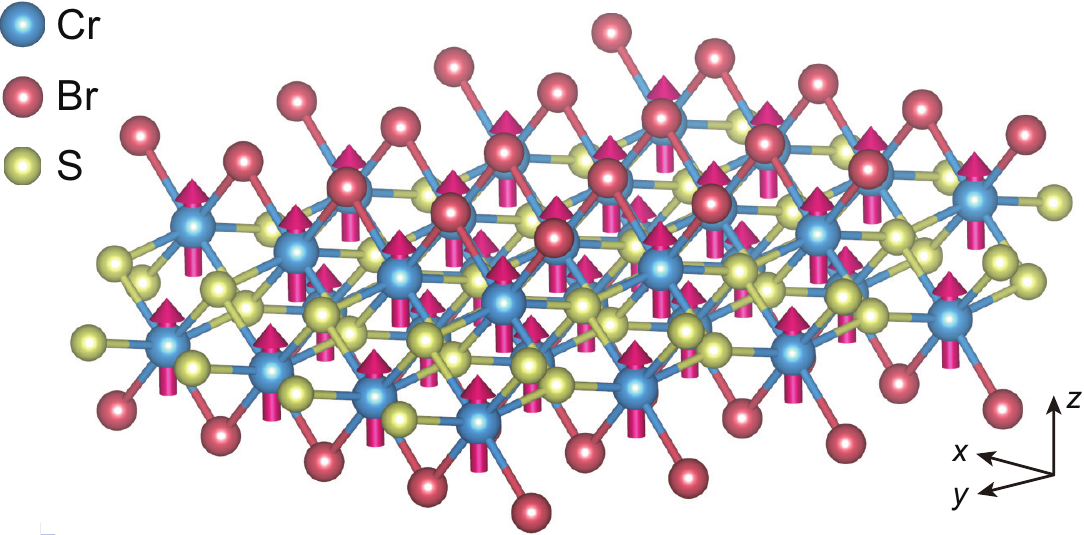}
	\caption{\label{fig:20} Illustration of crystal structure of 2D magnet CrSBr monolayer with out-of-plane magnetization.}
\end{figure}

In 2D magnets, the in-plane spin orientation could be tuned to out-of-plane direction by external perpendicular magnetic field. As shown in Fig. 20, here we consider the out-of-plane magnetization along \textit{z} axis ($\bm{B}=B\bm{\hat{e}_{z}}$). For an isotropic system the dielectric tensor is expressed as:
\begin{equation}
	\bm{\varepsilon}(\omega,\bm{B}) =
	\begin{pmatrix}
		\varepsilon_{xx}(\omega,\bm{B}) & \varepsilon_{xy}(\omega,\bm{B}) & 0 \\
		-\varepsilon_{xy}(\omega,\bm{B}) & \varepsilon_{xx}(\omega,\bm{B}) & 0 \\
		0 & 0 & \varepsilon_{zz}(\omega,\bm{B}) \\
	\end{pmatrix}.
\end{equation} The Fresnel equation for the propagation of electromagnetic wave is given by:
\begin{equation}
	[n^{2}\bm{I}-\bm{\varepsilon}-\bm{n}:\bm{n}]\cdot\bm{E}=0.
\end{equation}
For normal incidence, the complex refractory index \textbf{n} is:
\begin{equation}
	\bm{n}=  \frac{c\bm{k}}{\omega}=\frac{ck}{\omega}\bm{\hat{e}_{z}}.
\end{equation}
By solving the above Fresnel equations, we 
get the normal modes as the $\sigma^+$ and $\sigma^-$ for circularly polarized plane waves, with distinct refractive indices:
\begin{equation}
	n^{2}_{\pm}(\omega,B\bm{\hat{e}_{z}})=\varepsilon_{xx}(\omega,B\bm{\hat{e}_{z}})\pm i\varepsilon_{xy}(\omega,B\bm{\hat{e}_{z}}),
\end{equation}
For an anisotropic system, the distinct refractive indices are:
\begin{equation}
	n^{2}_{\pm}(\omega,B\bm{\hat{e}_{z}})=\varepsilon_{0}(\omega,B\bm{\hat{e}_{z}})\pm i\varepsilon_{xy}(\omega,B\bm{\hat{e}_{z}}),
\end{equation}
\begin{equation}
	\varepsilon_{0}(\omega,B\bm{\hat{e}_{z}})= \frac{\varepsilon_{xx}(\omega,B\bm{\hat{e}_{z}})+\varepsilon_{yy}(\omega,B\bm{\hat{e}_{z}})}{2}.
\end{equation}

To mathematically describe how an electromagnetic wave interacts with such stratified and anisotropic media, we adopt a 4x4 formalism involving the in-plane components of both
electric $(E_{x}, E_{y})$ and magnetic fields $(B_{x}, B_{y})$. We consider a monolayer material
magnetized along the $+z$ direction, and the second layer is
vacuum. Within each layer ($l$ = 0, 1, 2), we choose the four
eigenmodes of light as follows:$\,\bm{\hat{e}^{(l)}_{1}}=\bm{\hat{e}^{(l)}_{2}}=\bm{\hat{e}_{+}}, \,\bm{\hat{e}^{(l)}_{3}}=\bm{\hat{e}^{(l)}_{4}}=\bm{\hat{e}_{-}}$, with the corresponding refractive indices:$\,n^{(l)}_{1}=-n^{(l)}_{2}=n^{(l)}_{+},\,n^{(l)}_{3}=-n^{(l)}_{4}=n^{(l)}_{-}$. The electric and magnetic fields of
light in the first and second layers are given by:
\begin{equation}
	\bm{E}^{(l)}=\sum_{j=1}^{4}E^{(l)}_{0j}\bm{\hat{e}}^{(l)}_{j}\exp \{i[k^{(l)}_{j}(z-z_{l-1})-\omega t] \},
\end{equation}
and
\begin{equation}
	c\bm{B}^{(l)}=\sum_{j=1}^{4}E^{(l)}_{0j}\bm{b}^{(l)}_{j}\exp \{i[k^{(l)}_{j}(z-z_{l-1})-\omega t] \},
\end{equation}
with $\bm{b}^{(l)}_{j} = n^{(l)}_{j}\bm{\hat{e}}_{z} \times \bm{\hat{e}}^{(l)}_{j}$. The electric and magnetic fields of light in the zeroth layer are given by,
\begin{equation}
	\bm{E}^{(0)}=\sum_{j=1}^{4}E^{(0)}_{0j}\bm{\hat{e}}^{(0)}_{j}\exp \{i[k^{(0)}_{j}(z-z_{0})-\omega t] \},
\end{equation}
and
\begin{equation}
	c\bm{B}^{(0)}=\sum_{j=1}^{4}E^{(0)}_{0j}\bm{b}^{(0)}_{j}\exp \{i[k^{(0)}_{j}(z-z_{0})-\omega t] \}.
\end{equation}
The requirement of the continuity of the tangential field
components at the interfaces connects the field amplitudes $E^{(l)}_{0j}$ between two neighboring layers. The dynamical matrix within each layer
is given by a block-diagonal form:
\begin{equation}
	\begin{aligned}
		\bm{D}^{(l)} &=
		\begin{pmatrix}
			\bm{\hat{e}}^{(l)}_{1} \cdot \bm{\hat{e}}_{+}^* & \bm{\hat{e}}^{(l)}_{2}\cdot \bm{\hat{e}}_{+}^* & \bm{\hat{e}}^{(l)}_{3}\cdot \bm{\hat{e}}_{+}^* &\bm{\hat{e}}^{(l)}_{4}\cdot \bm{\hat{e}}_{+}^*\\
			\bm{\hat{b}}^{(l)}_{1}\cdot \bm{\hat{e}}_{+}^* & \bm{\hat{b}}^{(l)}_{2}\cdot \bm{\hat{e}}_{+}^* & \bm{\hat{b}}^{(l)}_{3}\cdot \bm{\hat{e}}_{+}^* &\bm{\hat{b}}^{(l)}_{4}\cdot \bm{\hat{e}}_{+}^*\\
			\bm{\hat{e}}^{(l)}_{1}\cdot \bm{\hat{e}}_{-}^* & \bm{\hat{e}}^{(l)}_{2}\cdot \bm{\hat{e}}_{-}^* & \bm{\hat{e}}^{(l)}_{3}\cdot \bm{\hat{e}}_{-}^* &\bm{\hat{e}}^{(l)}_{4}\cdot \bm{\hat{e}}_{-}^*\\
			\bm{\hat{b}}^{(l)}_{1}\cdot \bm{\hat{e}}_{-}^* & \bm{\hat{b}}^{(l)}_{2}\cdot \bm{\hat{e}}_{-}^* & \bm{\hat{b}}^{(l)}_{3}\cdot \bm{\hat{e}}_{-}^* &\bm{\hat{b}}^{(l)}_{4}\cdot \bm{\hat{e}}_{-}^*\\
		\end{pmatrix} \\
		&=
		\begin{pmatrix}
			1 & 1 & 0 &0\\
			-in^{(l)}_{+} & in^{(l)}_{+} & 0 &0\\
			0 & 0 & 1 &1\\
			0 & 0 & in^{(l)}_{-} & -in^{(l)}_{-}\\
		\end{pmatrix}.
	\end{aligned}
\end{equation}
The propagation matrix is defined as a diagonal matrix:
\begin{equation}
	\bm{P}^{(1)}=
	\begin{pmatrix}
		e^{i\delta_{+}} & 0 & 0 &0\\
		0 & e^{-i\delta_{+}} & 0 &0\\
		0 & 0 & e^{i\delta_{-}} &0\\
		0 & 0 & 0 &e^{-i\delta_{-}}\\
	\end{pmatrix},
\end{equation}
where $\delta_{+}=\frac{\omega}{c}n_{+}$, $\delta_{-}=\frac{\omega}{c}n_{-}$, and $d$ is the thickness of layer 1, i.e. monolayer CrSBr. 

In this two-interface setup, $\bm{E}^{(0)}_{0}$
and $\bm{E}^{(2)}_{0}$ are related by the transfer matrix $\bm{M}_{c}$ in the basis of
linearly polarized lights:
\begin{equation}
	\bm{E}^{(2)}_{0}=\bm{M}_{c}\bm{E}^{(0)}_{0}=[\bm{D}^{(2)}]^{-1}\bm{D}^{(1)}\bm{P}^{(1)}[\bm{D}^{(1)}]^{-1}\bm{D}^{(0)}\bm{E}^{(0)}_{0}.
\end{equation}
$\bm{M}_{c}$ has a simple block-diagonal form:
\begin{equation}
	\bm{M}_{c}=
	\begin{pmatrix}
		\bm{M}_{+} & 0 \\
		0 & \bm{M}_{-} \\
	\end{pmatrix},
\end{equation}
\begin{equation}
	\bm{M}_{+}= \frac{1}{t^{+}_{21}t^{+}_{10}}
	\begin{pmatrix}
		e^{i\delta_{+}}+e^{-i\delta_{+}}r^{+}_{21}r^{+}_{10} & e^{i\delta_{+}}r^{+}_{10}+e^{-i\delta_{+}}r^{+}_{21} \\
		e^{i\delta_{+}}r^{+}_{21}+e^{-i\delta_{+}}r^{+}_{10} & e^{i\delta_{+}}r^{+}_{21}r^{+}_{10}+e^{-i\delta_{+}} \\
	\end{pmatrix},
\end{equation}
where $r_{mn}$ and $t_{mn}$ are the Fresnel coefficients of the interface
from the \textit{m}th layer to the \textit{n}th layer. And $\bm{M}_{-}$ has the same form.
In the left and right layers, we adopt a basis transformation from the circularly polarized light ($\bm{\hat{e}}_{+}$,$\bm{\hat{e}}_{-}$) with $\bm{\hat{e}}_{+}=-\frac{1}{\sqrt{2}}(\bm{\hat{e}}_{x}+i\bm{\hat{e}}_{y})$ and $\bm{\hat{e}}_{-}=\frac{1}{\sqrt{2}}(\bm{\hat{e}}_{x}-i\bm{\hat{e}}_{y})$ to the light ($\bm{\hat{e}}_{x}$,$\bm{\hat{e}}_{y}$). $\bm{\hat{e}}^{(l)}_{1}=\bm{\hat{e}}^{(l)}_{2}=\bm{\hat{e}}_{x}$, $\bm{\hat{e}}^{(l)}_{3}=\bm{\hat{e}}^{(l)}_{4}=\bm{\hat{e}}_{y}$, and $n^{(l)}_{1}=-n^{(l)}_{2}=n^{(l)}_{3}=-n^{(l)}_{4}=n^{(l)}$, for $l$ $=$ $0$, $2$. This new basis of linearly
polarized plane waves is denoted as $\{x\rightarrow,x\leftarrow,y\rightarrow,y\leftarrow\}$. In this basis of linearly polarized lights, the electric field amplitudes in the left and right layer are related by transfer matrix $M$:
\begin{equation}
	\begin{pmatrix}
		E^{(2)}_{0x\rightarrow} \\
		E^{(2)}_{0x\leftarrow} \\
		E^{(2)}_{0y\rightarrow} \\
		E^{(2)}_{0y\leftarrow} \\
	\end{pmatrix}
	=\bm{M}
	\begin{pmatrix}
		E^{(0)}_{0x\rightarrow} \\
		E^{(0)}_{0x\leftarrow} \\
		E^{(0)}_{0y\rightarrow} \\
		E^{(0)}_{0y\leftarrow} \\
	\end{pmatrix}
\end{equation}
\begin{equation}
	\bm{M}=\frac{1}{2}
	\begin{pmatrix}
		\bm{M}_{+}+\bm{M}_{-} & -i(\bm{M}_{+}-\bm{M}_{-}) \\
		i(\bm{M}_{+}-\bm{M}_{-}) & \bm{M}_{+}+\bm{M}_{-} \\
	\end{pmatrix}.
\end{equation}

Because the electric field of incident light is along the direction
of magnetization ($+z$), $E^{(2)}_{0y\leftarrow}=0$. In addition, there are no reflecting lights from the
\textit{zero}th medium (semi-infinite substrate) to CrSBr, which means $E^{(0)}_{0x\rightarrow}=E^{(0)}_{0y\rightarrow}=0$. With these two conditions, by calculating Eq. G16, we can get:
\begin{equation}
	t_{ss}\equiv\frac{E^{(0)}_{0x\leftarrow}}{E^{(2)}_{0x\leftarrow}}=\frac{\bm{M}_{44}}{\bm{M}_{22}\bm{M}_{44}-\bm{M}_{24}\bm{M}_{42}}
\end{equation}
\begin{equation}		t_{sp}\equiv\frac{E^{(0)}_{0y\leftarrow}}{E^{(2)}_{0x\leftarrow}}=\frac{-\bm{M}_{42}}{\bm{M}_{22}\bm{M}_{44}-\bm{M}_{24}\bm{M}_{42}}
\end{equation}
\begin{equation}
	r_{ss}\equiv\frac{E^{(0)}_{0x\rightarrow}}{E^{(2)}_{0x\leftarrow}}=\frac{\bm{M}_{12}\bm{M}_{44}-\bm{M}_{14}\bm{M}_{42}}{\bm{M}_{22}\bm{M}_{44}-\bm{M}_{24}\bm{M}_{42}}
\end{equation}
\begin{equation}
	r_{sp}\equiv\frac{E^{(0)}_{0y\rightarrow}}{E^{(2)}_{0x\leftarrow}}=\frac{\bm{M}_{32}\bm{M}_{44}-\bm{M}_{34}\bm{M}_{42}}{\bm{M}_{22}\bm{M}_{44}-\bm{M}_{24}\bm{M}_{42}}
\end{equation}
The Kerr signals for the reflected (right-moving)
electric field $E^{(2)}_{\rightarrow}=E^{(2)}_{0x\rightarrow}\bm{\hat{e}}_{x}+E^{(2)}_{0y\rightarrow}\bm{\hat{e}}_{y}$ are expressed as:
\begin{equation}
	\tan2\theta_{\rm{K}}= \frac{2|\frac{r_{sp}}{r_{ss}}|\cos(\arg\frac{r_{sp}}{r_{ss}})}{1-|\frac{r_{sp}}{r_{ss}}|^{2}},
\end{equation}
\begin{equation}
	\sin2\eta_{\rm{K}}= \frac{2|\frac{r_{sp}}{r_{ss}}|\sin(\arg\frac{r_{sp}}{r_{ss}})}{1+|\frac{r_{sp}}{r_{ss}}|^{2}},
\end{equation}
Similarly, the Faraday signals for the transmitted (left-moving)
electric field
$E^{(0)}_{\leftarrow}=E^{(0)}_{0x\leftarrow}\bm{\hat{e}}_{x}+E^{(0)}_{0y\leftarrow}\bm{\hat{e}}_{y}$ are expressed as:
\begin{equation}
	\tan2\theta_{\rm{F}}= \frac{2|\frac{t_{sp}}{t_{ss}}|\cos(\arg\frac{t_{sp}}{t_{ss}})}{1-|\frac{t_{sp}}{t_{ss}}|^{2}},
\end{equation}
\begin{equation}
	\sin2\eta_{\rm{F}}= \frac{2|\frac{t_{sp}}{t_{ss}}|\sin(\arg\frac{t_{sp}}{t_{ss}})}{1+|\frac{t_{sp}}{t_{ss}}|^{2}}.
\end{equation}

Based on the above derivation, in Fig. 21, we have shown rotation angles and ellipticities of Kerr and Faraday effects considering electron-hole interaction. The maximal rotation angles of Kerr and Faraday effects are 1.1$^\circ$ and 0.6$^\circ$, respectively, which are larger than 0.9$^\circ$ and 0.3$^\circ$ in CrI$_3$ \cite{wu-natcomm} and 0.3$^\circ$ and 0.2$^\circ$ in CrBr$_3$ \cite{wu-prmater}. Compared with Fig. 22, both the amplitude and the position of the spectrum are modified significantly for excitonic effect. This verifies the fact that it is excitonic effect that enhances the Kerr and Faraday effects in CrSBr monolayer.

\begin{figure}[H]
	\centering
	\includegraphics[width=0.45\textwidth]{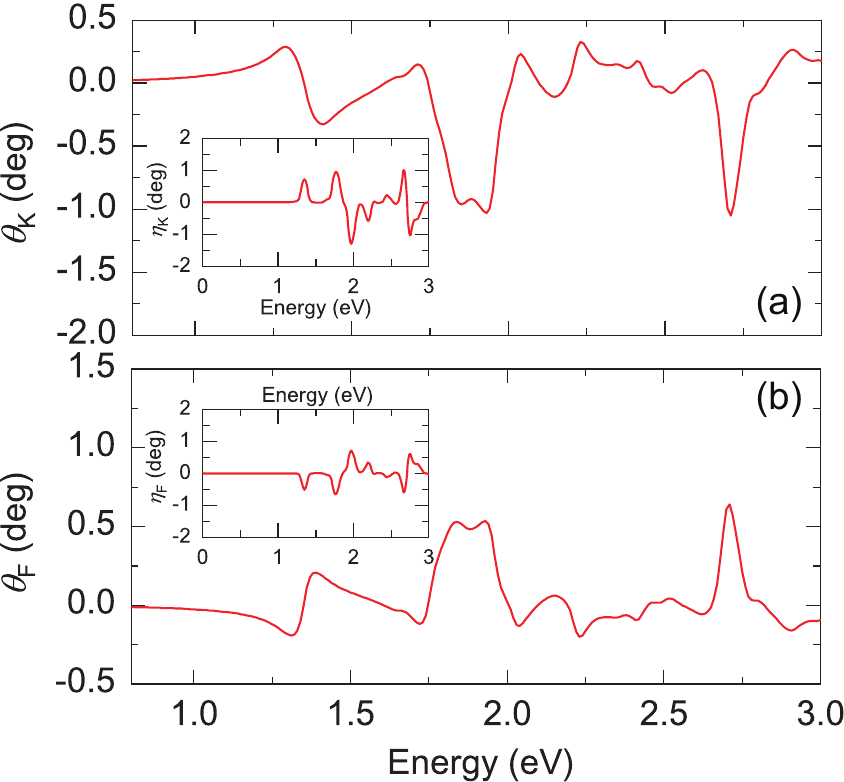}
	\caption{\label{fig:21} Magneto-optical Kerr and Faraday effects: Rotation angles of (a) reflected and (b) transmitted light in CrSBr monolayer for the state with spin orientation along $z$ axis considering electron-hole interactions. Insets are the corresponding ellipticities.}
\end{figure}

\begin{figure}[H]
	\centering
	\includegraphics[width=0.45\textwidth]{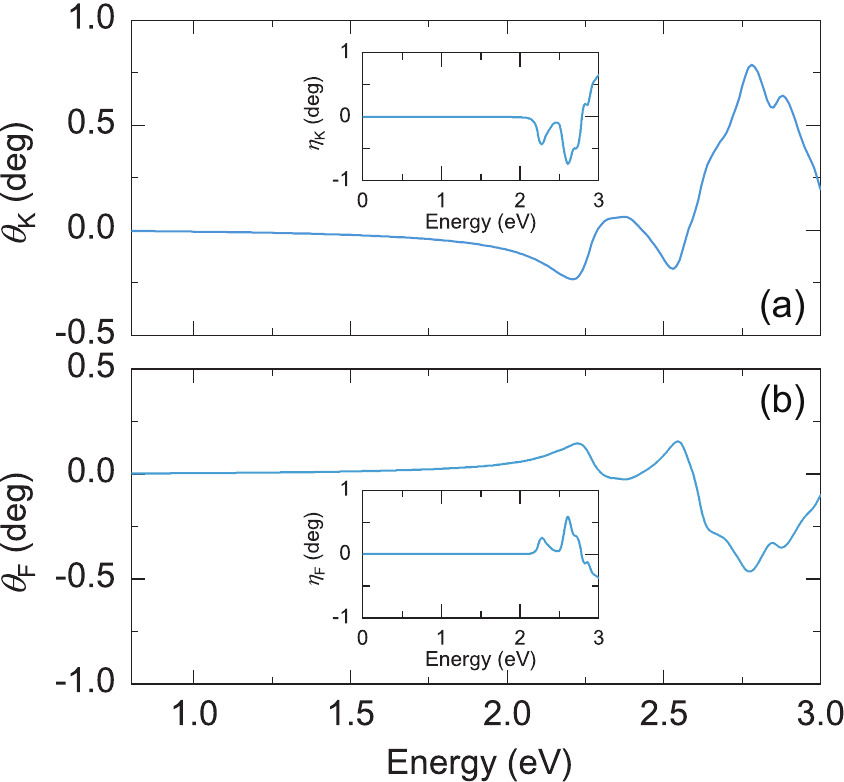}
	\caption{\label{fig:22} Magneto-optical Kerr and Faraday effects: Rotation angles of (a) reflected and (b) transmitted light in CrSBr monolayer for the state with spin orientation along $z$ axis without the inclusion of electron-hole interactions. Insets are the corresponding ellipticities.}
\end{figure}

\section{$G_0W_0$-BSE CALCULATIONS OF OPTICAL PROPERTIES IN BLACK PHOSPHORENE}

In Fig. 23, we have shown the anisotropic optical absorption spectrum and exciton states (within
$G_0W_0$-BSE framework) for black phosphorene. We use the
norm-conserving PBE pseudopotentials with a plane-wave cutoff of
80 Ry. For the $G_0W_0$ part, we use a coarse \emph{k} grid of
$18\times12\times1$, 500 empty bands (5 valence bands), and the dielectric cutoff of 10 Ry. For the BSE part,
a fine \emph{k} grid of $72\times48\times 1$ is used. A Gaussian
smearing with a broadening constant of 30 meV is used in optical
absorption spectrum. The number of bands for optical transitions
is 4 for both valence and conduction bands, which is
sufficient to cover the span of the visible light. The lattice constants are 3.303 {\AA} for \textit{a} and 4.625 {\AA} for \textit{b}. In Fig. 24, we show the linear dichroism for both reflected and transmitted light with rotation angle and ellipticity.

\begin{figure}[H]
	\centering
	\includegraphics[width=0.48\textwidth]{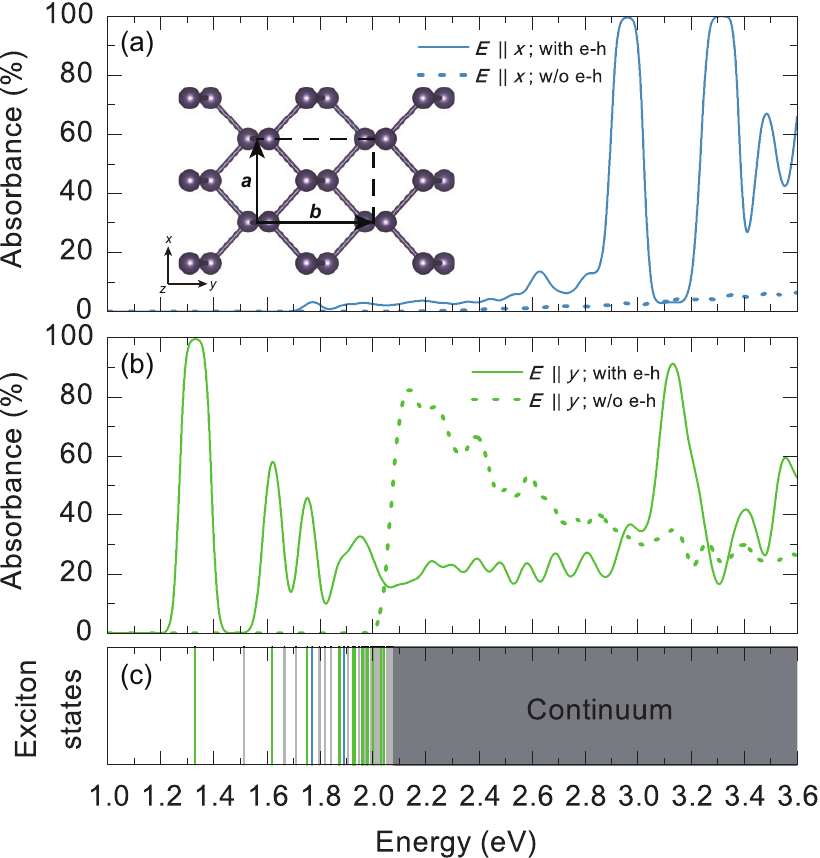}
	\caption{\label{fig:23} Anisotropic optical absorbance in black phosphorene monolayer for the electric polarization along \textit{x} (a) and \textit{y} (b) directions, respectively. (c) Exciton spectrum. Both optical spectra with and without the consideration of electron-hole interaction are presented. Inset of (a) is the crystal structure of black phosphorene and the gray lines are for dark excitons.}
\end{figure}

\begin{figure}[H]
	\centering
	\includegraphics[width=0.45\textwidth]{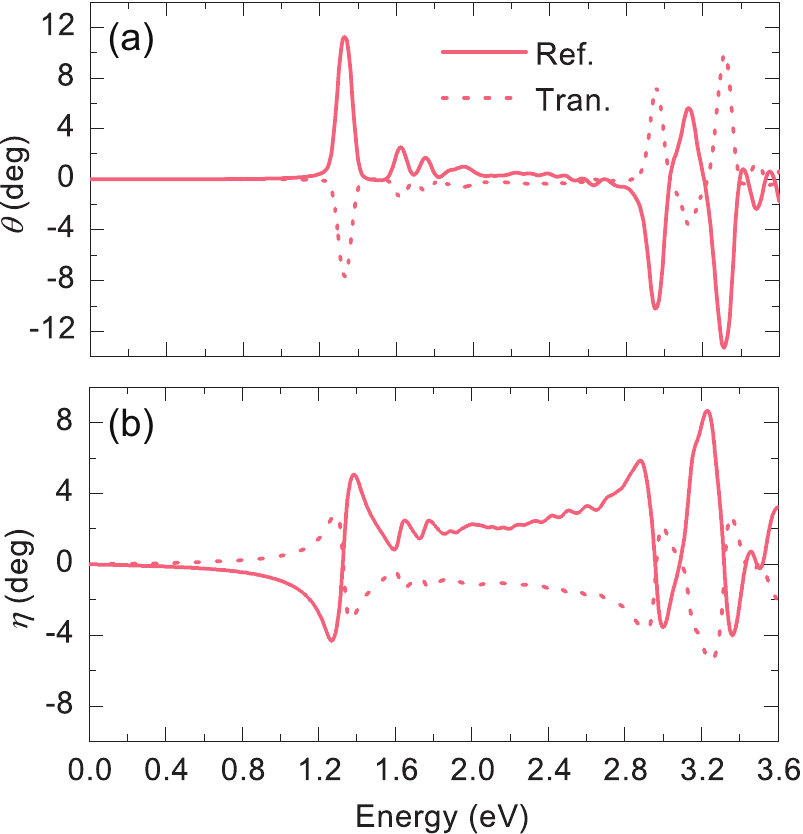}
	\caption{\label{fig:24} Rotation angles (a) and ellipticities (b) in black phosphorene.}
\end{figure}

\clearpage

\end{document}